\documentclass{aa}
\input epsf

\title{Reddenings and metallicities in the LMC and SMC from 
       Str{\"o}mgren CCD photometry
\thanks{Based on observations carried out with the Str{\"o}mgren Automatic
	Telescope (SAT) and the Danish 1.54 meter telescope at ESO, La Silla}
\thanks{The CCD $uvby$ photometry is available in electronic form at the CDS
        via anonymous ftp to 130.79.128.5 or via http://cdsweb.u-strasbg.fr/Abstract.html}
}
\author{S.S. Larsen, \inst{1,2}
        \and J.V. Clausen \inst{1}
	\and J. Storm \inst{3}}

\offprints{S.S. Larsen}

\institute{
Niels Bohr Institute for Astronomy, Physics and Geophysics,
Astronomical Observatory,
Juliane Maries Vej 30, DK-2100 Copenhagen {\O}, Denmark \\
	   email: jvc@astro.ku.dk
        \and UC Observatories / Lick Observatory, University of California,
	  Santa Cruz, CA 95064, USA \\
	  email: soeren@ucolick.org
        \and Astrophysikalisches Institut Potsdam,
	An der Sternwarte 16, D-14482 Potsdam, Germany \\
	email: jstorm@aip.de}
\begin{document}
\newcommand{\h}{\mbox{$^{\mbox{\small h}}$}}
\newcommand{\m}{\mbox{$^{\mbox{\small m}}$}}
\newcommand{\s}{\mbox{$^{\mbox{\small s}}$}}

\thesaurus{03(11.01.01; 11.09.4; 11.13.1; 11.16.1}


\date{Received 25 May 2000 / Accepted 11 October 2000}

\maketitle
\begin{abstract}
  The {\it individual} reddenings for B stars in two fields in the Small 
Magellanic Cloud (SMC) and two fields in the Large Magellanic Cloud (LMC) 
are determined by means of Str{\"o}mgren $uvby$ CCD photometry. In both 
LMC fields we find a foreground reddening of $E(B-V) = 0.085 \pm 0.02$, 
and for the SMC fields we find $E(B-V) = 0.070 \pm 0.02$. In addition to 
the foreground reddening we find contributions from reddening intrinsically 
in the Clouds up to $E(B-V) \sim 0.21$. The intrinsic contribution is 
not correlated with position within our $\sim6\arcmin\times4\farcm5$ CCD 
fields but varies in an essentially random way.
Unless the reddening is measured for a particular object, it will be 
uncertain by $\pm 0.035$ (best case, far from the central bars) to more than 
$\pm 0.10$ (close to the central bars).

  The Str{\"o}mgren $vby$ photometry has been used to derive metallicities
for GK giant stars in the observed fields. Adopting average reddenings
we obtain mean metallicities  which are consistent with those found from 
spectroscopic studies of F and G supergiants (Westerlund \cite{west1997}),
but with a considerable scatter in the derived metallicities, from 
[Fe/H] $\approx -2.0$ to [Fe/H] $\ga 0$.  A significant fraction of the 
scatter is, however, due to reddening variations rather than being 
intrinsic. The possible existence of high metallicity stars should be 
investigated further using spectroscopic methods.



\end{abstract}

\section{Introduction}

  This paper describes an investigation of reddenings and metallicities 
in two fields in each of the Magellanic Clouds by means of Str{\"o}mgren 
CCD photometry. For the first time, the reddening
variations on very small scales ($\sim1$ pc, limited by the surface 
density of B stars) have been examined from Str{\"o}mgren photometry. 
A detailed analysis of the metallicities of GK giant 
stars is performed, with special emphasis on the problems that arise 
from an incomplete knowledge of the individual reddenings of the stars. 
The metallicities derived for GK giants from Str{\"o}mgren photometry are very 
sensitive to the assumed reddening, but because we have an estimate of the
scatter in the reddening distributions from the early-type stars we are 
able to give quantitative estimates of the uncertainty on metallicities
for individual GK giants.

  The disagreement in the literature on the reddening for Magellanic Cloud 
stars has been remarkably large. Part of the controversy probably arises from 
an insufficient distinction between foreground reddening and reddening
intrinsically in the Clouds.  
Grieve \& Madore (\cite{grie1986}) found that some stars in the \object{SMC} 
have reddenings as high as $E(B-V) = 0.20$ with an average of around 
$E(B-V) = 0.09$, while the maximum for the LMC is $E(B-V) = 0.3$ and the 
average is $E(B-V) = 0.1$. Grieve \& Madore (\cite{grie1986}) also noted 
that in the LMC the stars with high reddenings are predominantly concentrated 
in the neighbourhood of the 30 Doradus complex.  
In a review paper, Bessel (\cite{bess1991}) concluded that 
while it has been claimed that the reddening is as low as 
$E(B-V) = 0.01$ or $0.02$ for both Clouds, several investigations arrive at 
considerably higher reddenings, up to $E(B-V) = 0.1$ or even more.  
From studies of a (fairly small) section of the LMC, Harris et al. 
(\cite{harris1997})
determined a mean total $E(B-V)$ of 0.20 with a non-Gaussian tail to high 
values.
Oestreicher et al. (\cite{oest1995}) and 
Oestreicher \& Schmidt-Kaler (\cite{oest1996}) determined 
foreground reddenings and intrinsic reddenings for the \object{LMC}, 
and found that the 
foreground reddening generally varies from $E(B-V) = 0.0$ to $E(B-V) = 0.1$ 
across the face of the LMC, while reddenings up to $E(B-V) = 0.8$ were seen 
for some of the brightest A and B supergiants. However, the extreme reddenings 
were interpreted as being caused by circumstellar dust shells, an assumption 
which was supported by the fact that for the fainter stars only more modest 
reddenings, up to $E(B-V) = 0.20$, were detected. 
Schlegel et al. (\cite{schlegel1998}) report typical foreground reddenings
of $E(B-V) = 0.075$ (LMC) and 0.037 (SMC), respectively.
Using $UBVI$ photometry, Zaritsky~(\cite{zar99}) also found a dependence 
of reddening on spectral type, with stars hotter than $\sim 12000$ K having
visual extinctions up to several tenths of a magnitude larger than cooler
stars. Recently, Romaniello et al.~(\cite{rom00}) found a mean
$E(B-V)$ of 0.20 mag around SN 1987A with a scatter of 0.072 mag, claimed
to be at least twice the measurement errors.

  Many investigations have aimed at deriving metallicities for different
types of objects (HII regions, planetary nebulae, single stars) in the 
Magellanic Clouds. This is not the place to give an extensive 
review of all of them; we refer to Westerlund (\cite{west1990}, 
\cite{west1997}).
A general picture of a roughly exponential increase in metal abundance 
with time over the last 10 Gyr is often quoted for both galaxies. 
From studies of the surroundings of six LMC clusters, Dirsch et al. 
(\cite{dirsch2000}) have determined an age-metallicity relation which shows
a strong increase in the metallicity starting around 2--3 Gyrs ago.
For very young objects mean \mbox{[Fe/H]} values
are about $-0.2$ (LMC) and $-0.5$ (SMC), respectively, whereas ``canonical'' 
field star metallicities of around \mbox{[Fe/H] = $-0.3$} for the LMC and 
\mbox{[Fe/H] = $-0.65$} for the SMC are often used. 

  The Str{\"o}mgren $uvby\beta$ photometric system was originally designed 
for the study of BAF stars (Str{\"o}mgren \cite{strom1966}). It has, however, 
turned out to be very useful also for investigations of other types of stars.
Bond (\cite{bond1980}) demonstrated that metallicities can be derived for 
red giants by means of the Str{\"o}mgren $(b-y)_0$ and $m_0$ indices. In 
nearby galaxies, such as the Magellanic Clouds, the Str{\"o}mgren system 
thereby provides us with a potentially very powerful tool for studying the 
metallicities of large numbers of red field giant stars. 

Several {\it spectroscopic} studies of cool supergiants reach the
canonical metallicities and a scatter of about $\pm 0.2$ dex (e.g. 
Spite \& Spite 
\cite{spi1987}; Olszewski et al.~\cite{ols1991}; Hill \cite{hill1999}), 
whereas attempts to derive metallicities for red field giants from 
Str{\"o}mgren photometry (Hilker et al. \cite{hilk1995} (HRG95);
Grebel \& Richtler \cite{greb1992} (GR92); Dirsch et al. \cite{dirsch2000}) 
have failed to reproduce them.  GR92 studied the neighbourhood of the 
young cluster NGC 330 in the SMC and found [Fe/H] = $-1.26$ for the cluster 
itself, whereas the surrounding field stars had an average metallicity of 
[Fe/H]$ = -0.74$ with a large spread. HRG95 found 
[Fe/H] = $-0.93 \pm 0.16$ for NGC 330 and metallicities in the
range [Fe/H] = $-2.0$ to $-0.2$ for the field stars, with a peak 
at [Fe/H] $ \sim -1.0$.  For the red giants 
in the LMC cluster NGC 1866, HRG95 found a metallicity of
[Fe/H] = $-0.43 \pm 0.18$, while the field stars also here showed 
a large variation in metallicity with a peak at \mbox{[Fe/H] = $-0.7$}. 
Dirsch et al. (\cite{dirsch2000}) generally find metallicities lower
than $-0.3$ for six relatively young LMC clusters and their surrounding 
fields.

  In this paper we first derive reddenings for early-type stars in each
of the four fields using new CCD Str{\"o}mgren photometry. Metallicities
are then derived for red giant stars in the same four fields, also
using Str{\"o}mgren photometry. The uncertainties on the derived
metallicities are then discussed, with particular emphasis on the 
scatter resulting from an inaccurate knowledge of the reddenings of
individual red giant stars.

\section{Observations and reductions}

\begin{table}
\caption[]{\label{tab:fields}The observed fields. For comparison, the
positions for the optical centers of the bars of SMC and LMC are included. 
}
\begin{flushleft}
\begin{tabular}{lll} \hline
Field ID & $\alpha$(2000.0) & $\delta$(2000.0) \\ \hline
\object{LMC}     & $05\h24\m    $ & $-69\degr44\arcmin         $ \\  
\object{HV982}   & $05\h29\m53\s$ & $-69\degr09\arcmin23\arcsec$ \\  
\object{HV12578} & $05\h21\m32\s$ & $-66\degr21\arcmin15\arcsec$ \\ 
\object{SMC}     & $00\h53\m    $ & $-72\degr49\arcmin         $ \\  
\object{HV1433}  & $00\h47\m11\s$ & $-73\degr41\arcmin18\arcsec$ \\  
\object{HV11284} & $00\h49\m43\s$ & $-72\degr51\arcmin10\arcsec$ \\  
\hline
\end{tabular}
\end{flushleft}
\end{table}

\begin{table}
\caption{
Data for the combined frames. The first two columns are self-explanatory,
the third column gives the number of frames used in the combined frames,
the total integration time (in minutes) is listed in the fourth column, and 
the last column gives the resulting seeing measured on the combined frames.
\label{tab:cmb}
}
\begin{flushleft}
\begin{tabular}{lllll} \hline
Field & filter & $N$ & $T_{exp}$ & seeing \\ \hline
HV982 & $u$ & 67 & 640 & 1\farcs5 \\
     & $v$ & 87 & 250 & 1\farcs3 \\
     & $b$ & 62 & 180 & 1\farcs4 \\
     & $y$ & 199 & 500 & 1\farcs3 \\ \hline
HV12578 & $u$ & 28 & 560 & 1\farcs5 \\ 
     & $v$ & 40 & 200 & 1\farcs4 \\
     & $b$ & 28 & 140 & 1\farcs4 \\
     & $y$ & 100 & 500 & 1\farcs3 \\ \hline
HV1433 & $u$ & 16 & 290 & 1\farcs6 \\
     & $v$ & 24 & 160 & 1\farcs4 \\
     & $b$ & 21 & 140 & 1\farcs4 \\
     & $y$ & 53 & 272 & 1\farcs4 \\ \hline
HV11284 & $u$ & 21 & 420 & 1\farcs5 \\
     & $v$ & 29 & 145 & 1\farcs4 \\
     & $b$ & 31 & 155 & 1\farcs4 \\
     & $y$ & 84 & 420 & 1\farcs4 \\ \hline
\end{tabular}
\end{flushleft}
\end{table}

The observations were carried out during several observing runs in
1992 -- 1995 with the Danish 1.54 m telescope 
at ESO, La Silla, equipped with a direct camera and CCD \#28 (a 
$1024 \times 1024$ Tek device). This combination of telescope and CCD
yields a field size of about $6\farcm5 \times 6\farcm5$, although the
effective field size was somewhat smaller, typically 
$6\arcmin\times4\farcm5$, because we combined each image
from many individual frames that were not perfectly aligned. 
The image scale is 0\farcs38 / pixel.  Four fields centered on
eclipsing variable stars were observed, two in the SMC and two in the 
LMC (Table~\ref{tab:fields}). The HV12578 field is located close to the 
region E of the {\em ESO Key Programme on Coordinated Investigations of 
Selected Regions in the Magellanic Clouds} (de Boer et al.~\cite{deboer1989}) 
in the northern part of the LMC, whereas the HV982 field is 
located close to the Key Programme region F, in the northern outskirt of 
the LMC Bar, close to the 30 Doradus complex. The HV1433 and HV11284 fields 
are located on each side of the Key Programme B region, in the southern 
part of the SMC.

  The observations were originally obtained for a study of eclipsing 
binary B stars (Clausen et al., in prep.). Therefore, the data set 
consists of a large number of short individual exposures (typically 5 minutes 
in $vby$ and 20 minutes in $u$). Initial reductions, essentially flatfielding
and bias subtraction, were carried out using standard 
IRAF\footnote{
IRAF is distributed by the National Optical Astronomical Observatories,
which are operated by the Association of Universities for Research in
Astronomy, Inc. under contract with the National Science Foundation.
}
tools.
For each field and each filter, exposures
with seeing better than $1\farcs7$ and with sky background levels lower
than a certain threshold were selected, and these exposures were
then summed using the IRAF task {\bf imcombine}. Hence, equivalent 
integration times of typically 3--5 hours in $y,b$ and $v$, and $5-10$ 
hours in $u$
were obtained (see Table~\ref{tab:cmb}). The combination of many short 
exposures offered the advantage that cosmic ray events could be effectively 
eliminated, using the {\bf crreject} option in {\bf imcombine}.

  The calibration to the standard $uvby$ system was based on observations of 
secondary standard stars defined by photoelectric observations with the 
Str{\"o}mgren Automatic Telescope (SAT) at ESO, La Silla 
(see Clausen et al. \cite{clausen1997}). The SAT 
observations were carried out simultaneously with the 1.54 m observations,
and also provided extinction coefficients for the calibration of the CCD 
photometry.  The rms difference between the transformed 
CCD magnitudes and SAT magnitudes of the secondary standard stars was less 
than 0.01 in $(b-y)$ and around 0.015 in $c_1$. Although the individual
CCD exposures were obtained during several observing runs, the instrumental
system was stable enough that we did not find
any reason to use separate transformations for the different runs, except
for zero-point differences. This indicates that the properties of the filters
and the CCD
are quite stable, while the zero-point differences may be due to changes
in the reflectivity of telescope mirrors, gain settings for the CCD camera
electronics etc.

For the CCD photometry we used DAOPHOT II (Stetson \cite{stetson1987}) 
running within IRAF. The internal accuracy of the CCD photometry 
was typically around 0.01 mag in $(b-y)$ and 0.03 mag in $c_1$ and $m_1$
for stars brighter than $V\sim19$, while we find 
that systematic errors due to the determination of
zero-points for the PSF photometry relative to the aperture photometry of the
standard stars are on the order of 0.01 mag in all filters. 

  It is a notorious
problem to get the zero-points of the crowded-field PSF photometry relative
to the standard aperture photometry right, in particular in a case like ours
where there are no bright, isolated stars present in the fields. We approached
the problem as follows: First, a set of individual exposures in the four
Str{\"o}mgren filters were selected for each field. Photometry was obtained 
on all stars in the selected frames following the usual DAOPHOT procedure, and 
then all stars except for a few bright ones were subtracted using the task
{\bf substar} in the DAOPHOT package.  The few stars that were not 
subtracted were then measured using standard aperture photometry,
thus providing a set of possible ``tertiary standard stars'' in each frame. 
The final tertiary standard stars were selected from careful curves of
growth analyses; see Larsen (\cite{soeren1996}) for further details.
The PSF photometry of the combined frames was then tied into the aperture 
photometry system using the tertiary standards in each frame.

  Finally Be stars were identified from $H\alpha$ exposures and only 
between 5 and 39 of these stars were found in the various fields. It was 
also tested that foreground stars cause no serious problems; see Larsen 
(\cite{soeren1996}) for details.

\section{Reddenings}

\subsection{Reddenings from Str{\"o}mgren photometry}

  The classical way to derive reddenings from Str{\"o}mgren photometry
is to use the $\beta$ index and $(b-y)$ which are very well correlated for 
F-type stars (Crawford \cite{craw1975}).  The $\beta$ index is unaffected by
reddening, so the reddening can be determined by comparing the observed
$(b-y)$ to the value expected from the standard $\beta$-$(b-y)$ relation.
However, in the Magellanic Clouds the F-type main sequence stars are much 
too faint to be reached by our photometry, and there are additional problems 
with the $\beta$ photometry and its calibration because of the 
non-vanishing radial velocities of the clouds (Knude \cite{knud1993}).

  However, the Str{\"o}mgren system also provides a way to determine
reddenings for the more luminous early-type stars (A0 and earlier,
Str{\"o}mgren \cite{strom1966}). Instead of $(b-y)$ and $\beta$, the reddening
for early-type stars is derived from the $(b-y)$ and $c_1$ indices, 
following the ``standard iterative procedure'' described e.g. by Crawford 
et al. (\cite{craw1970}). Both $(b-y)$ and $c_1$ are temperature indicators for stars 
of spectral type earlier than A0, and for such stars the de-reddened 
indices $(b-y)_0$ and $c_0$ are related to each other through a standard 
relation in much the same way as are $(b-y)$ and $\beta$ in the case of 
F stars.  
The calibration of the $(b-y)_0 - c_0$ standard relation was empirically 
established by Crawford (\cite{craw1978}), and agrees well with Str{\"o}mgren indices 
based on model atmosphere calculations for early-type stars (Grebel, private 
communication. See also Sec.~\ref{sec:fgred} and Fig.~\ref{fig:byc1_synt}). 
The basic assumption is then that any observed deviation 
from the standard relation is caused by reddening. 

  The reddening vector in the $(b-y),c_1$ diagram is 
nearly horizontal so the reddening of a star can, to the first order, 
be estimated simply as the difference between the observed $(b-y)$ and the 
intrinsic value $(b-y)_0$ corresponding to the observed $c_1$ according to 
the standard relation:
\[
  E(b-y) \approx (b-y) - (b-y)_0 (c_1)
\]
With this initial estimate of the reddening, the 
$c_1$ index can be corrected and a new reddening can be derived. Usually only 
one or two iterations are required.

 Due to the smaller wavelength difference between the $b$ and $y$ filters 
compared to the Johnson $B$ and $V$ filters, $E(B-V)$ is somewhat larger than 
$E(b-y)$, as can be shown from the standard law for interstellar extinction 
(Savage \& Mathis \cite{sav1979}). More specifically, $E(B-V)$ and $E(b-y)$ are related 
through the following expression:

\begin{equation}
  E(B-V) = 1.4 \times E(b-y)
\end{equation}

  A very convenient point regarding the de-reddening of stars earlier than
spectral type A0 is that the $\beta$ index is not required. 
However, because 
the $c_1$ index involves the $u$ filter $(c_1 = (u-v) - (v-b))$, it has 
until recently been a time-consuming process to acquire sufficiently deep 
photometry for stars in the Magellanic Clouds with available CCD detectors.

\subsection{Selection of B stars}

\begin{figure}
\epsfxsize=8cm
\epsfbox{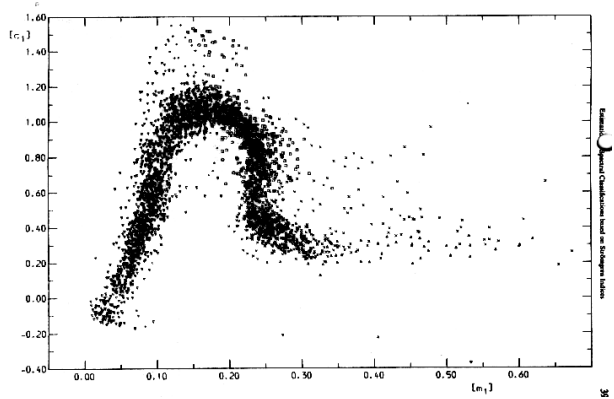}
\caption[]{\label{fig:m1c1_olsen}$[m_1],[c_1]$ diagram of
bright galactic stars (Olsen, \cite{olsen1979})}
\end{figure}

\begin{figure}
\epsfxsize=8cm
\epsfbox{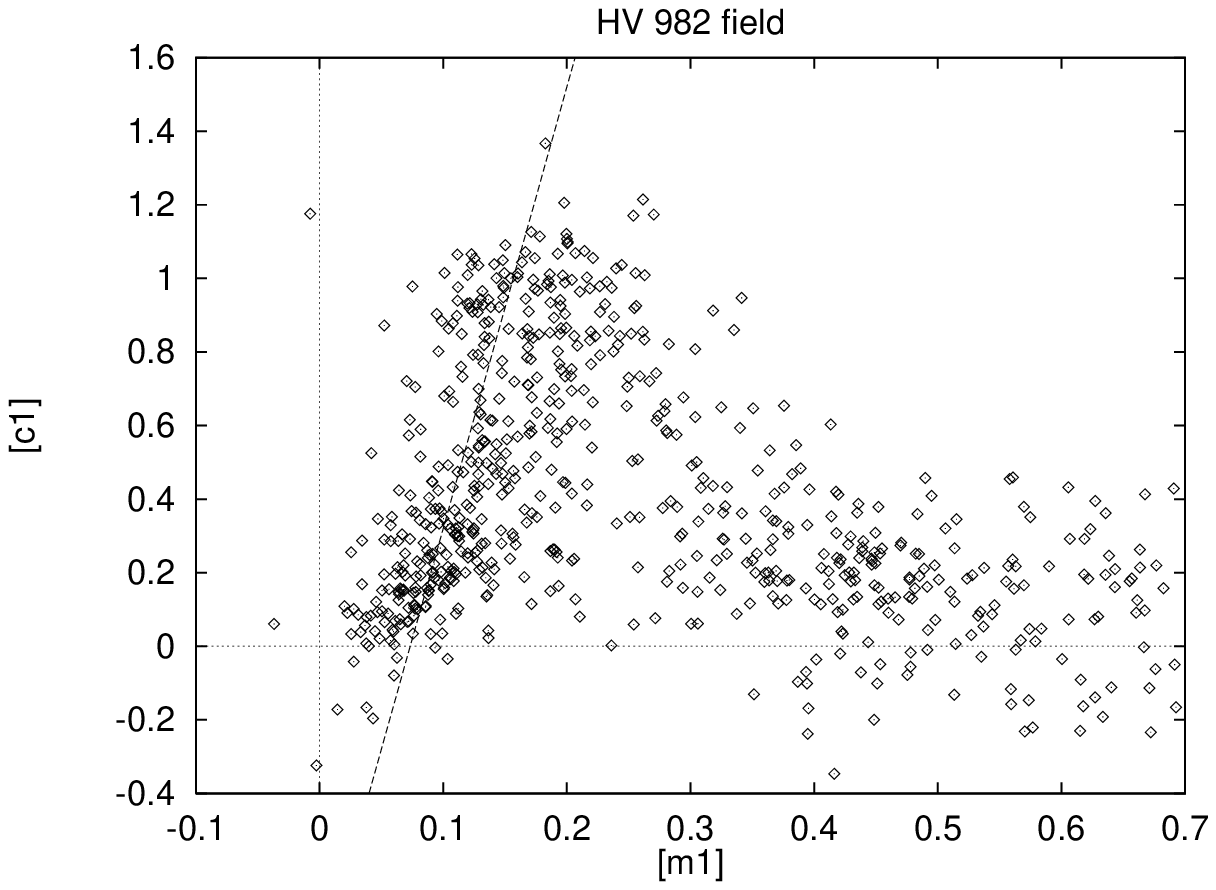}
\epsfxsize=8cm
\epsfbox{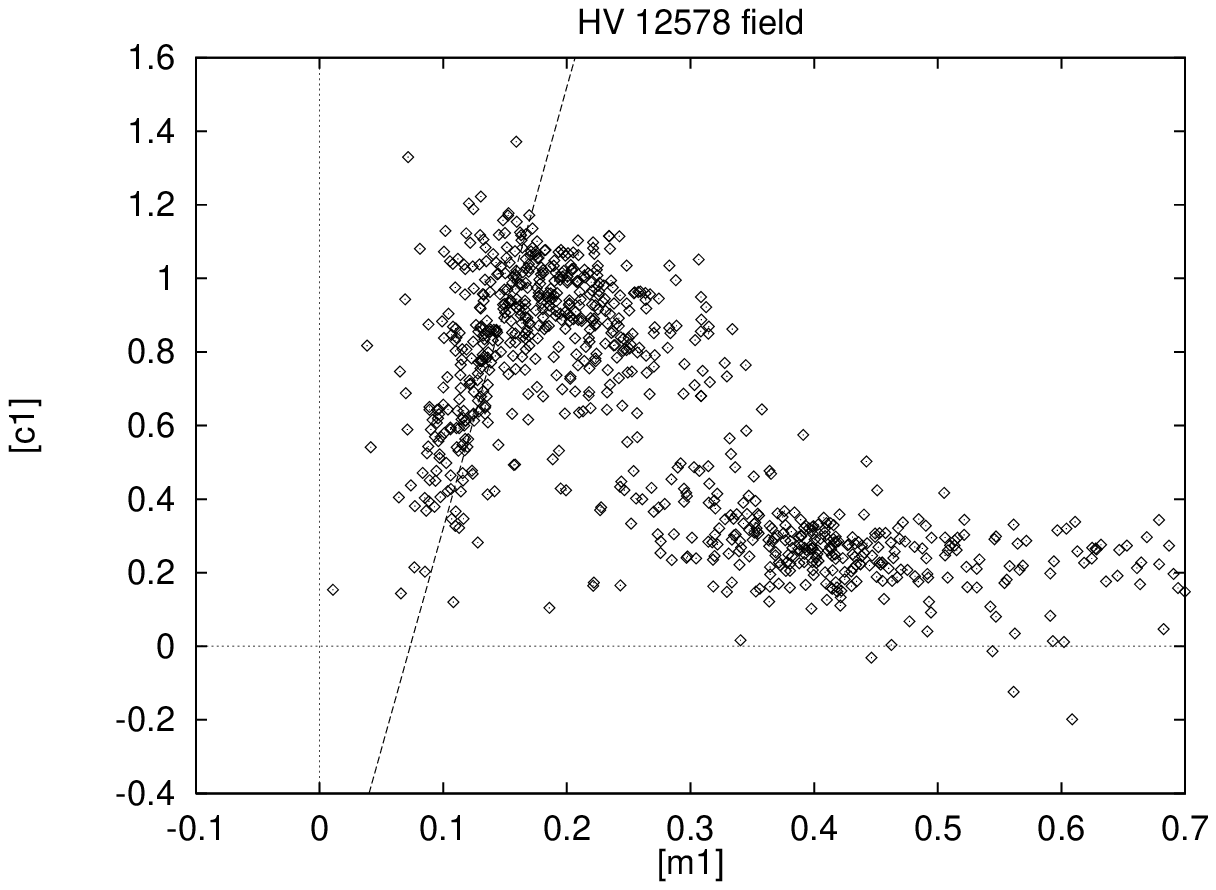}
\epsfxsize=8cm
\epsfbox{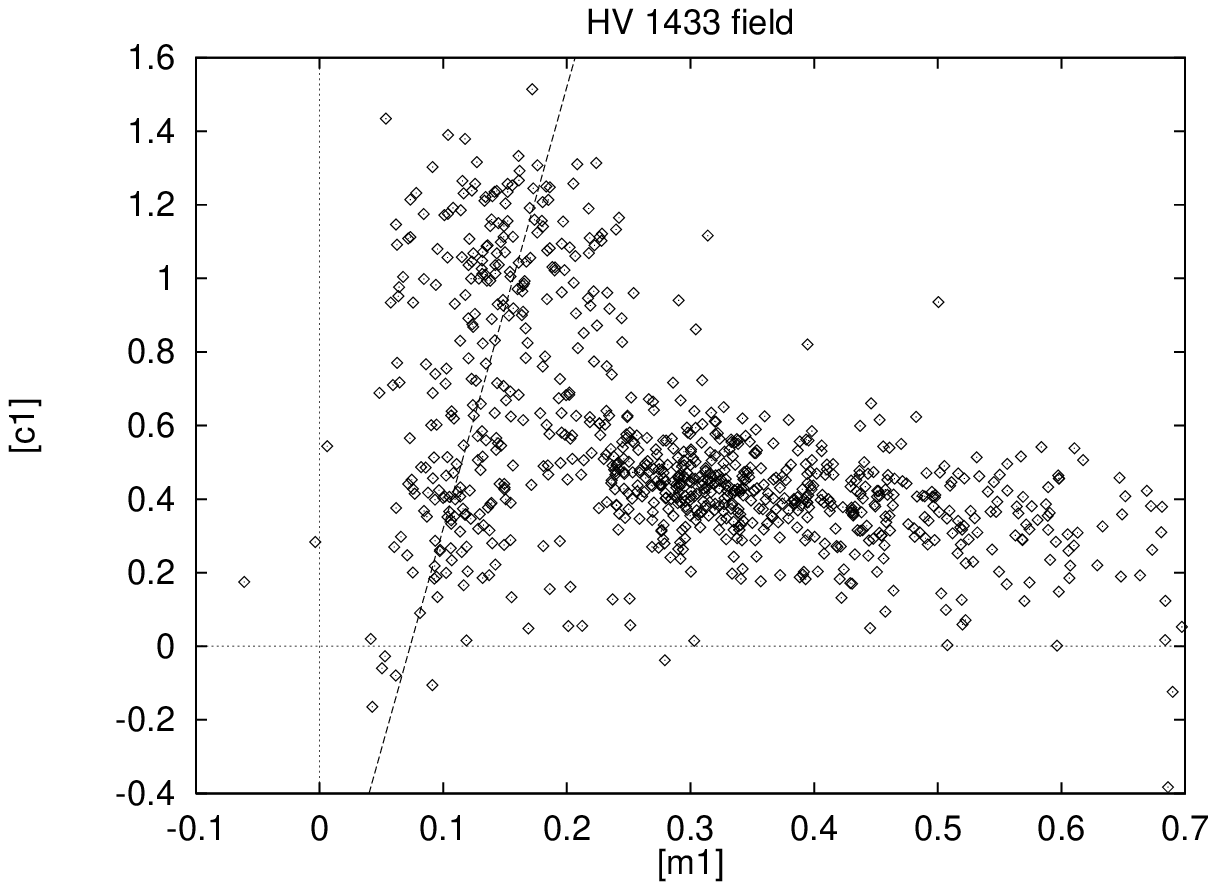}
\epsfxsize=8cm
\epsfbox{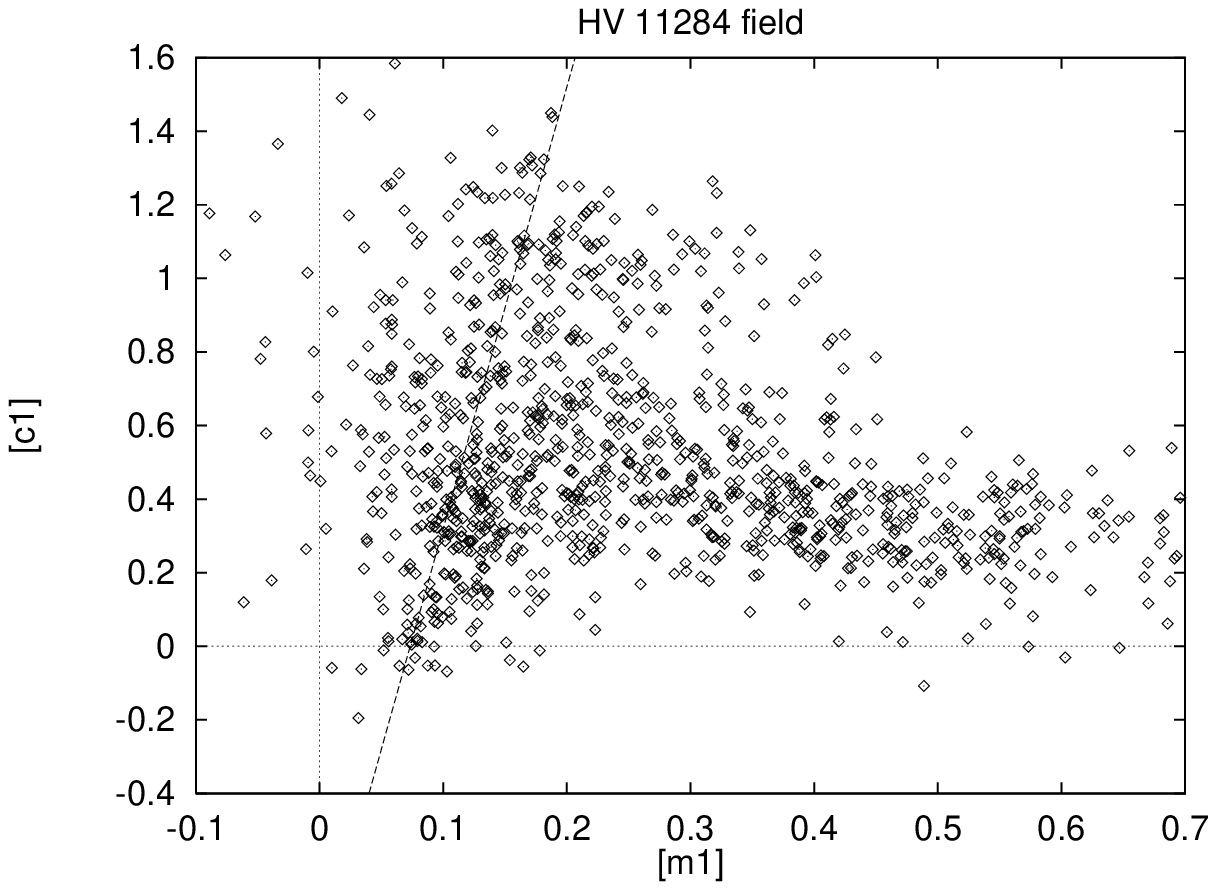}
\caption[]{\label{fig:m1c1}$[m_1],[c_1]$ diagrams for the four LMC/SMC fields.}
\end{figure}

  In principle, the selection of early-type stars is most effectively carried 
out in the Str{\"o}mgren system using the two reddening free indices 
\[
[m_1] = m_1 + 0.30 \, (b-y)
\]
and
\[
[c_1] = c_1 - 0.20 \, (b-y)
\]
According to Str{\"o}mgren (\cite{strom1966}) and Olsen (\cite{olsen1979}) 
the $[m_1],[c_1]$ diagram should resemble the one in Fig. \ref{fig:m1c1_olsen}. 
The early-type stars are located in the left part of the diagram, on the
part where $[c_1]$ is increasing as a function of $[m_1]$.
The lower, right border of the band containing the early-type stars is defined
quite well by a straight line given by the equation 
\begin{equation}
[c_1] = 12 \times [m_1]-0.88 
\label{eq:eholine}
\end{equation}
(E. H. Olsen, private communication).  The exact location of the line 
is only critical for $[c_1] > 0.9$, since no stars are found immediately 
to the right of it for lower $[c_1]$ values.

  The $[m_1],[c_1]$ diagrams for stars in the LMC/SMC fields with
${\rm err}(c_1) < 0.10$ mag (corresponding to 
$V \la 19.5$) are shown in Fig.~\ref{fig:m1c1}.  Obviously, they are 
not quite similar to the diagram shown in Fig. \ref{fig:m1c1_olsen}. 
The line separating early-type stars and stars of later types passes 
right through the distribution of points even for low $[c_1]$ values, 
and many stars are located inside the supposedly empty ``loop''. 

We have made several attempts at understanding this effect.  The 
observations and data reduction procedures have been carefully checked 
to ensure that the effect is not a manifestation 
of some error in the stacking procedure, the cosmic ray elimination process or
the use of DAOPHOT. This has been done by using the PSF fitting programme
DoPHOT (Schechter et al.~\cite{schecter1993}) on a set of single $uvby$ frames 
without removal of cosmic rays. Apart from the expected larger scatter, the
diagrams look qualitatively identical to the ones based on the full data
set.

\begin{figure*}
\epsfxsize=15cm
\epsfbox{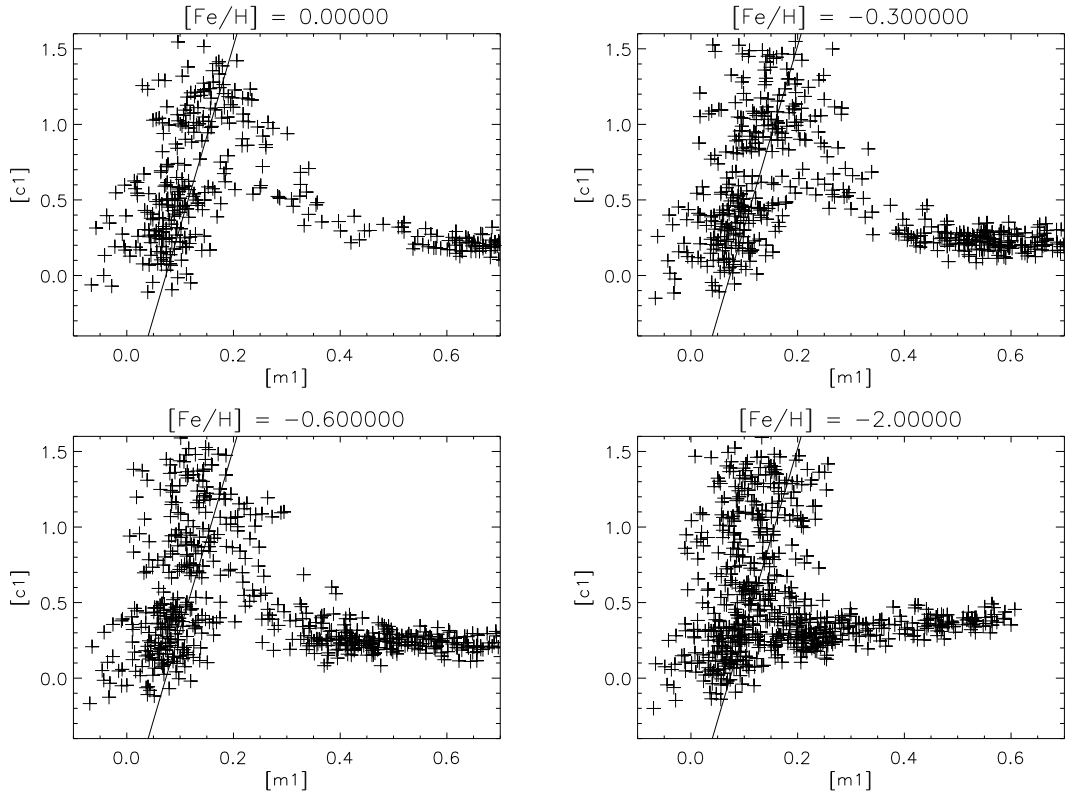}
\caption{\label{fig:m1c1_synt} Synthetic $[m_1],[c_1]$ diagrams based
on Padua stellar models and Kurucz model atmospheres. Gaussian
distributed random errors ($\sigma = 0.05$ mag) have been added to
the synthetic $[m_1]$ and $[c_1]$ values.
}
\end{figure*}

  Could the peculiar $[m_1],[c_1]$ diagrams result from the lower
metallicities in the Magellanic Clouds relative to local stars from which
Fig.~\ref{fig:m1c1_olsen} was derived? In order to test this we obtained
theoretical $[m_1],[c_1]$ colours for a set of stellar isochrones from
the Padua group (Girardi et al.~\cite{gir00}), transformed into
Str{\"o}mgren colours using 1997 versions of Kurucz model atmospheres 
(Kurucz~\cite{kur79}, \cite{kur92}).
In Fig.~\ref{fig:m1c1_synt} we show synthetic $[m_1],[c_1]$ diagrams
for four different metallicities ([Fe/H] = 0, $-0.3$, $-0.6$ and $-2.0$).
Each panel contains 5 isochrones corresponding to ages between
$10^7$ and $10^9$ years.  Only stars in the magnitude interval
$-1 < M_V < -4$, roughly corresponding to the range covered by our
data, have been included in the plot. Gaussian distributed random errors 
of $\sigma = 0.05$ mag, about similar to the typical errors in the 
observations, have been added to each axis (note that the 0.10 mag error 
limit in $c_1$ is an {\it upper} limit). The plots 
in Fig.~\ref{fig:m1c1_synt} do suggest a significant change in the 
appearance of the $[m_1],[c_1]$ diagrams for progressively lower 
metallicities in the sense that more late-type stars will move in from 
the right and fill the empty region of the diagram. This effect
is enhanced by the photometric scatter, although the early-type stars 
show only a very small systematic shift and for the HV982 field in 
particular, the theoretical $[m_1],[c_1]$ diagrams still do not provide 
a quite satisfactory match to the observations. Compared to 
Fig.~\ref{fig:m1c1_olsen}, Fig.~\ref{fig:m1c1} shows only few early-type 
stars {\em above} the expected location, whereas random photometric 
errors in Fig.~\ref{fig:m1c1_synt} cause equal amounts of stars to 
scatter upwards and downwards. However, the number of data points in 
different parts of the $[m_1],[c_1]$ diagram depends on the age distribution 
of the stars and unless this is taken into account, an exact match between
the observed and simulated distributions cannot be expected.

  We thus conclude that much of the apparently peculiar morphology of 
the observed $[m_1],[c_1]$ diagrams in the LMC and SMC could be due to 
a combination of lower metallicities and photometric errors.  In any case, 
for our purpose the $[m_1],[c_1]$ diagram clearly is not a practical tool
for selecting B stars.

  Therefore, we have eventually chosen the alternative approach of selecting 
B stars directly as stars brighter than V=19.0 in the LMC and V=19.4 in 
the SMC and bluer than $(b-y) = 0.2$.  The numbers of B stars in each field
selected in this way range between 102 (for the HV12578 field) and 322
(for the HV982 field).  The $(b-y)$ limit was chosen so that no confusion 
with red giants occurred, while nearly all main sequence stars would be 
included.  The magnitude limit corresponds to $M_V \sim 0$.
A potential problem is that evolved stars of later types than A0 
could enter our sample, but these would have relatively high $c_1$ values, 
and any significant contamination by stars later than type A0 should therefore
be visible in the \mbox{$(b-y),c_1$} diagram as an excess of stars with 
$c_1 \approx 1$. No such excess is observed.

\subsection{Foreground reddening}
\label{sec:fgred}

\begin{figure}
\epsfxsize=8cm
\epsfbox{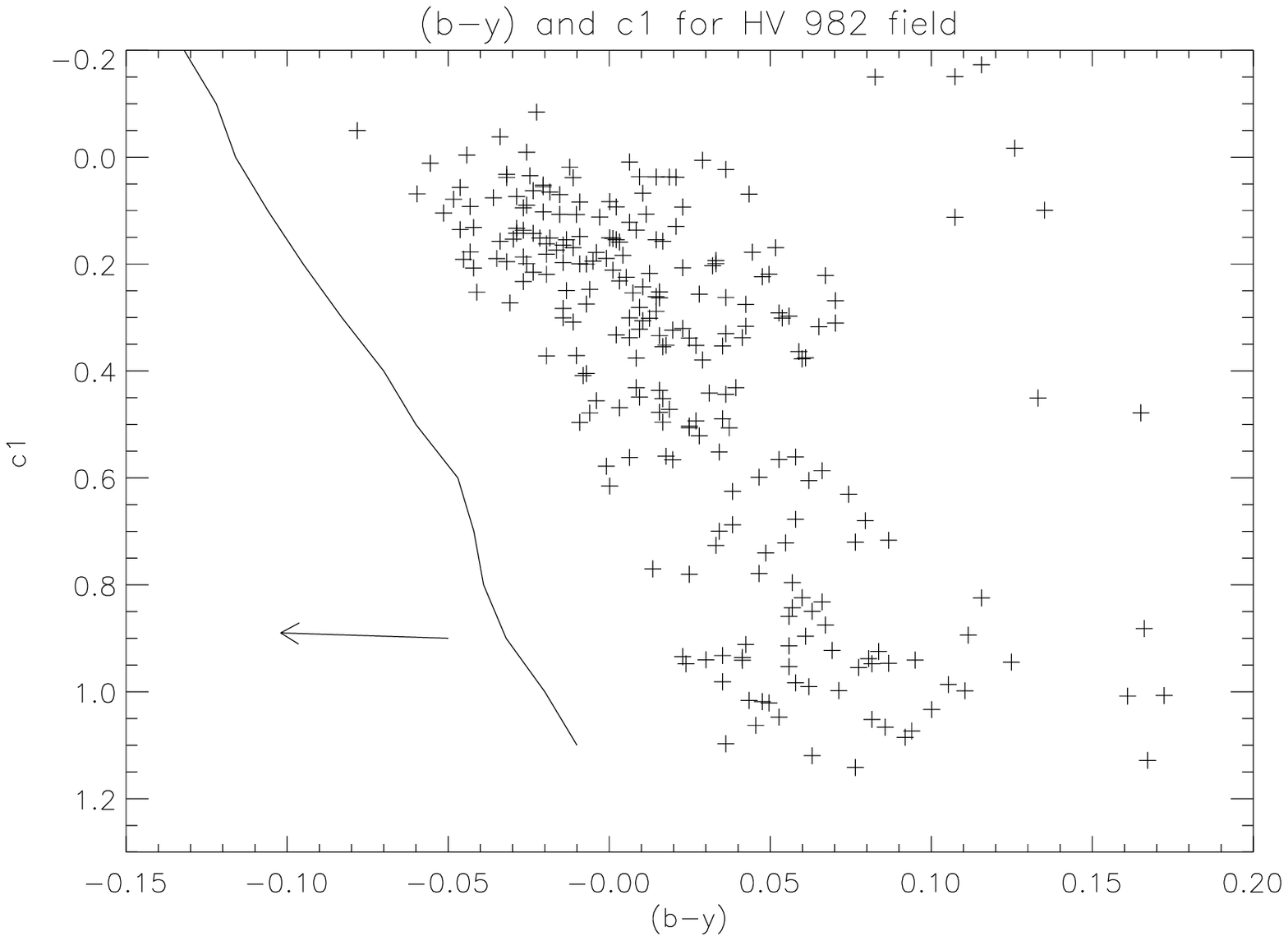}
\epsfxsize=8cm
\epsfbox{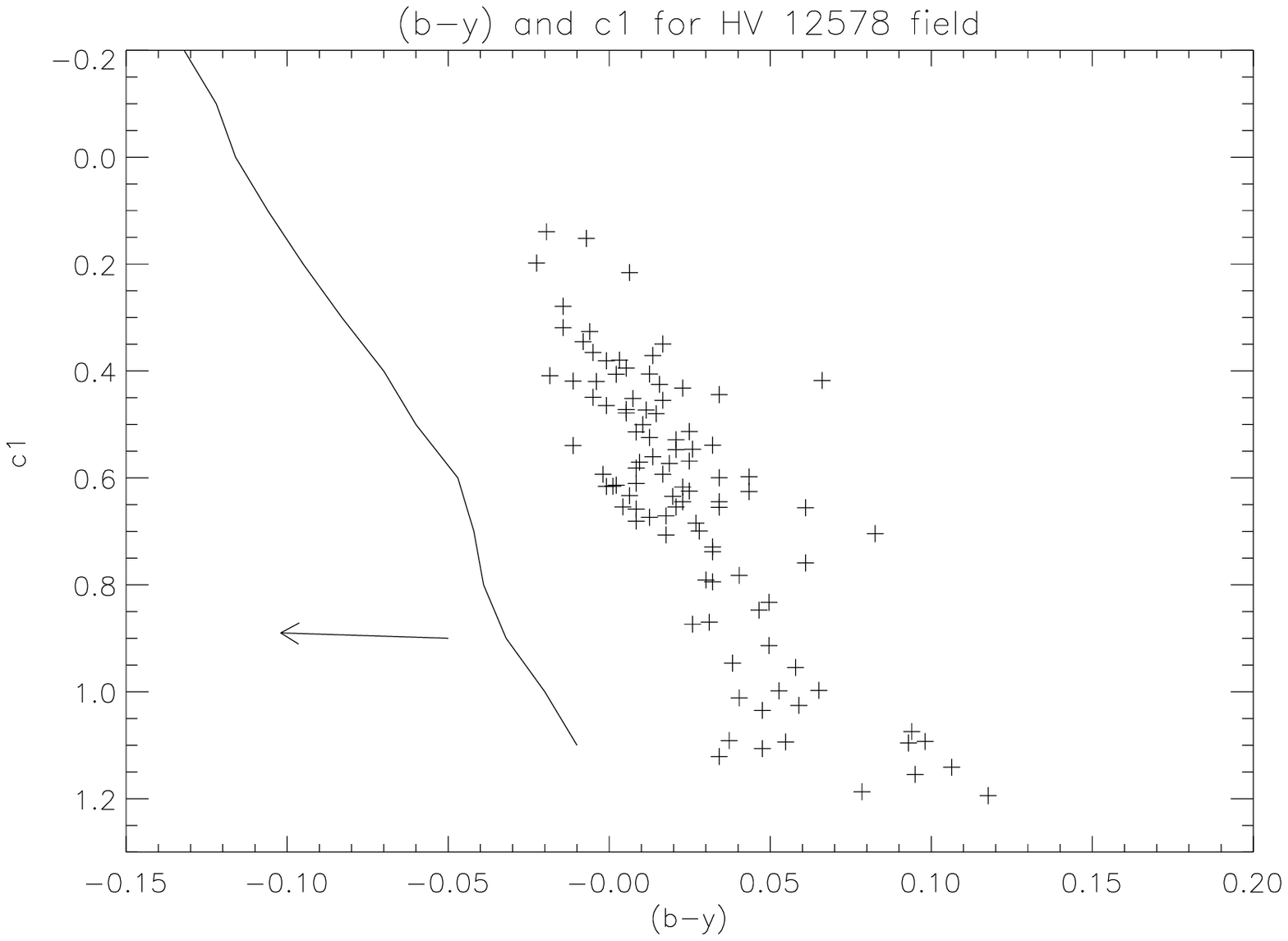}
\epsfxsize=8cm
\epsfbox{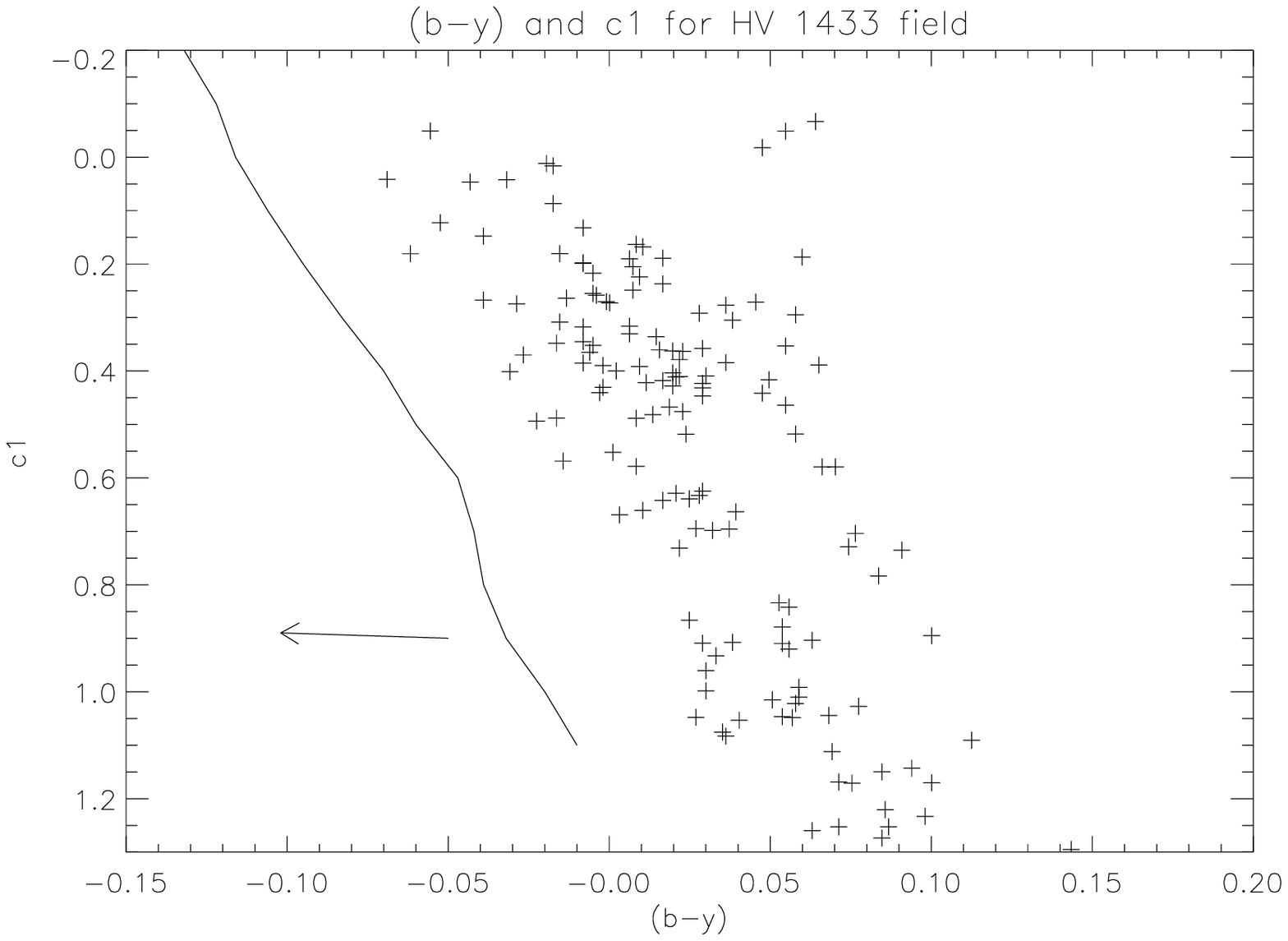}
\epsfxsize=8cm
\epsfbox{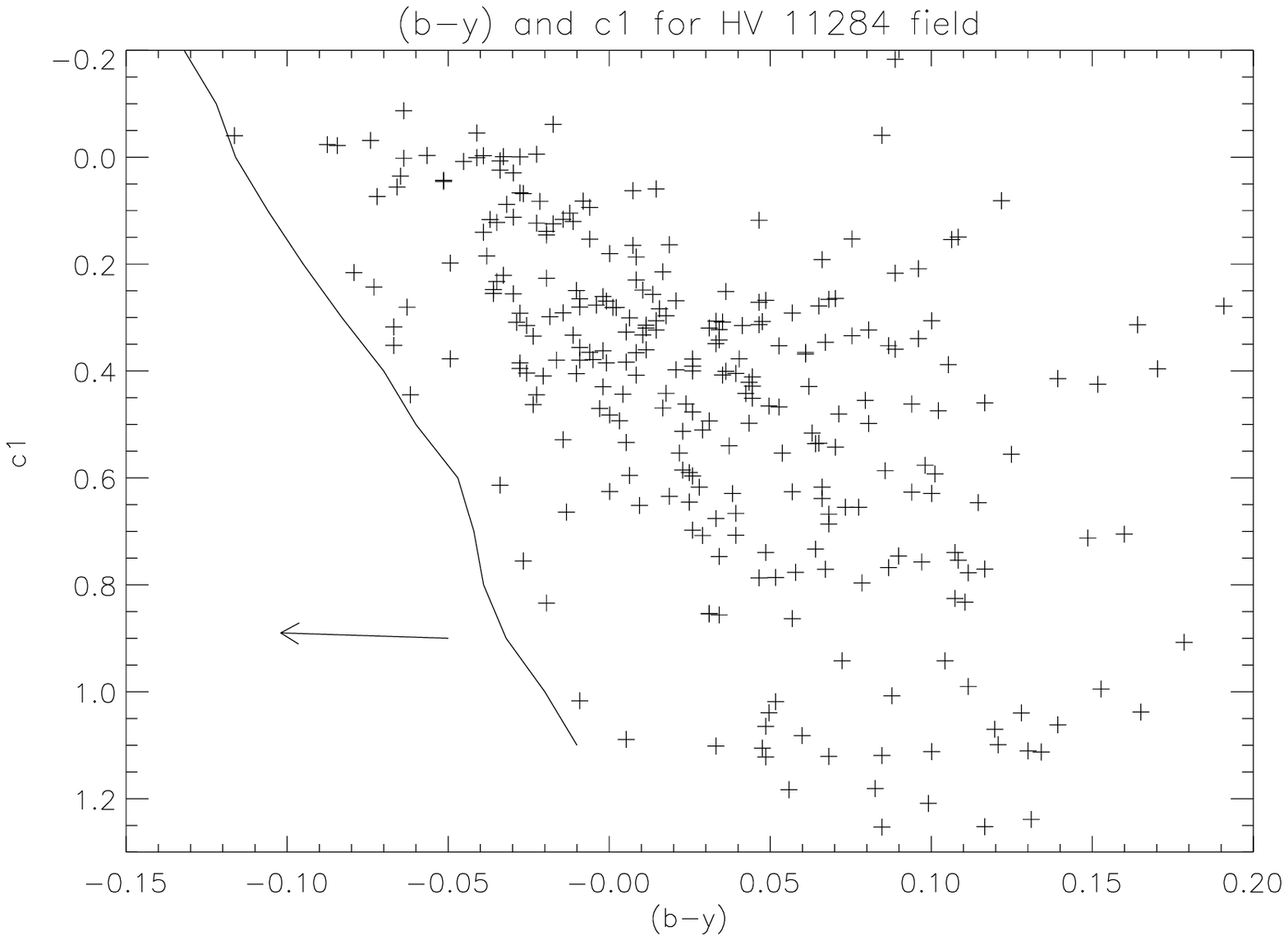}
\caption[]{\label{fig:byc1}$(b-y),c_1$ diagrams for B stars in the four 
fields. The arrow indicates the direction of the reddening vector.}
\end{figure}

\begin{figure}
\epsfxsize=8cm
\epsfbox{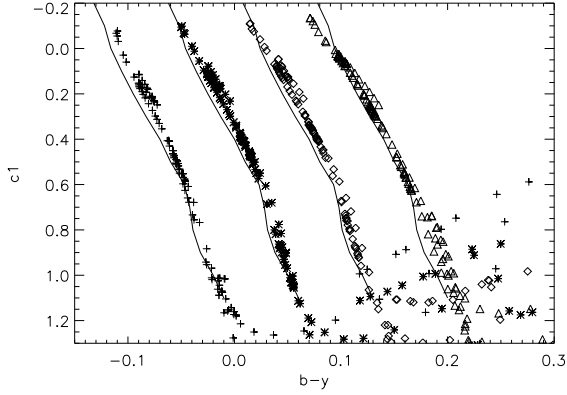}
\caption{\label{fig:byc1_synt} Synthetic $b-y,c_1$ diagrams based
on Padua stellar models and Kurucz model atmospheres. The plot shows
data for four metallicities, offset by steps of 0.07 mag in $(b-y)$.
Metallicities (left to right) are [Fe/H] = 0, $-0.3$, $-0.6$ and
$-2.0$.
}
\end{figure}

\begin{figure}
\epsfxsize=8cm
\epsfbox{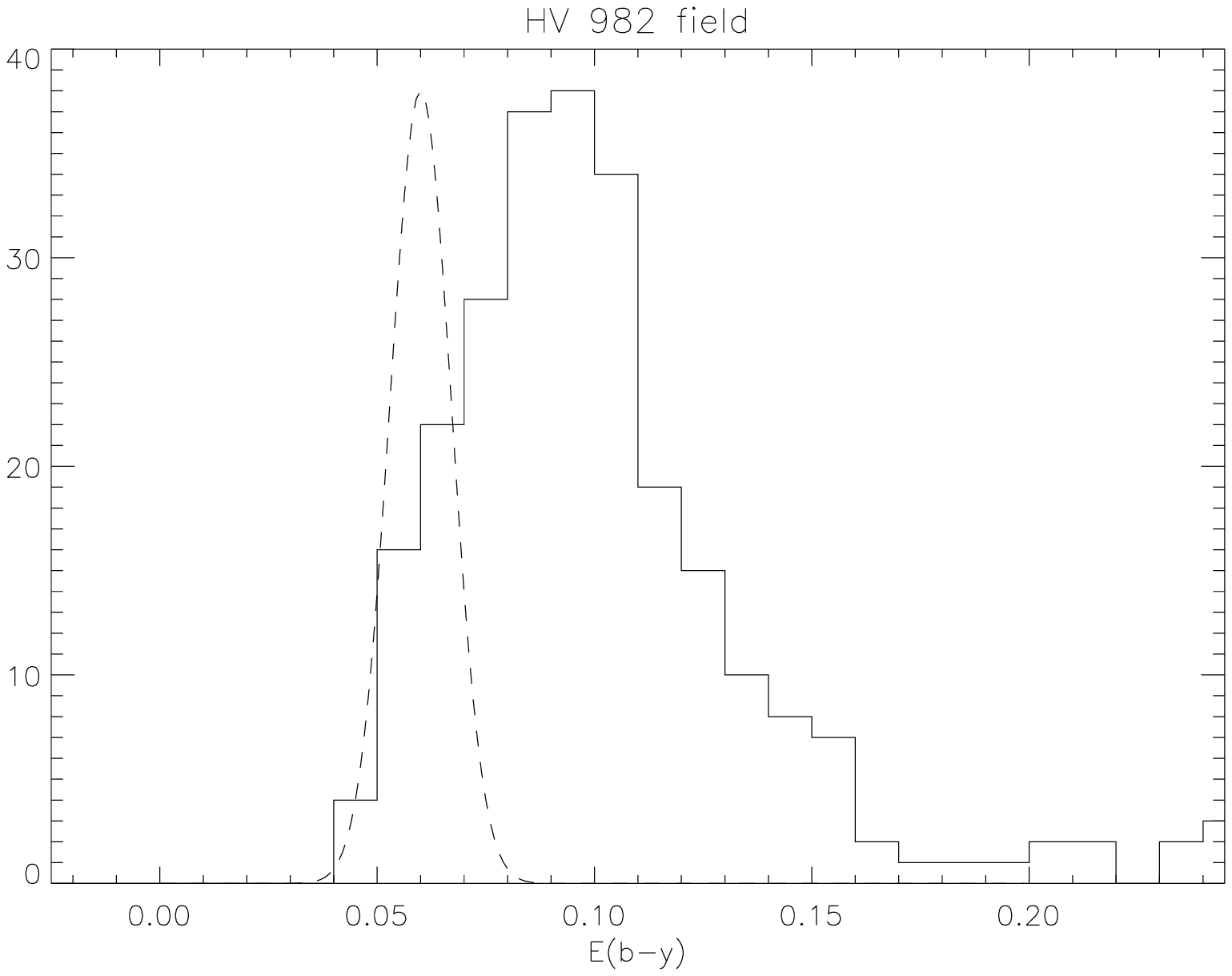}
\epsfxsize=8cm
\epsfbox{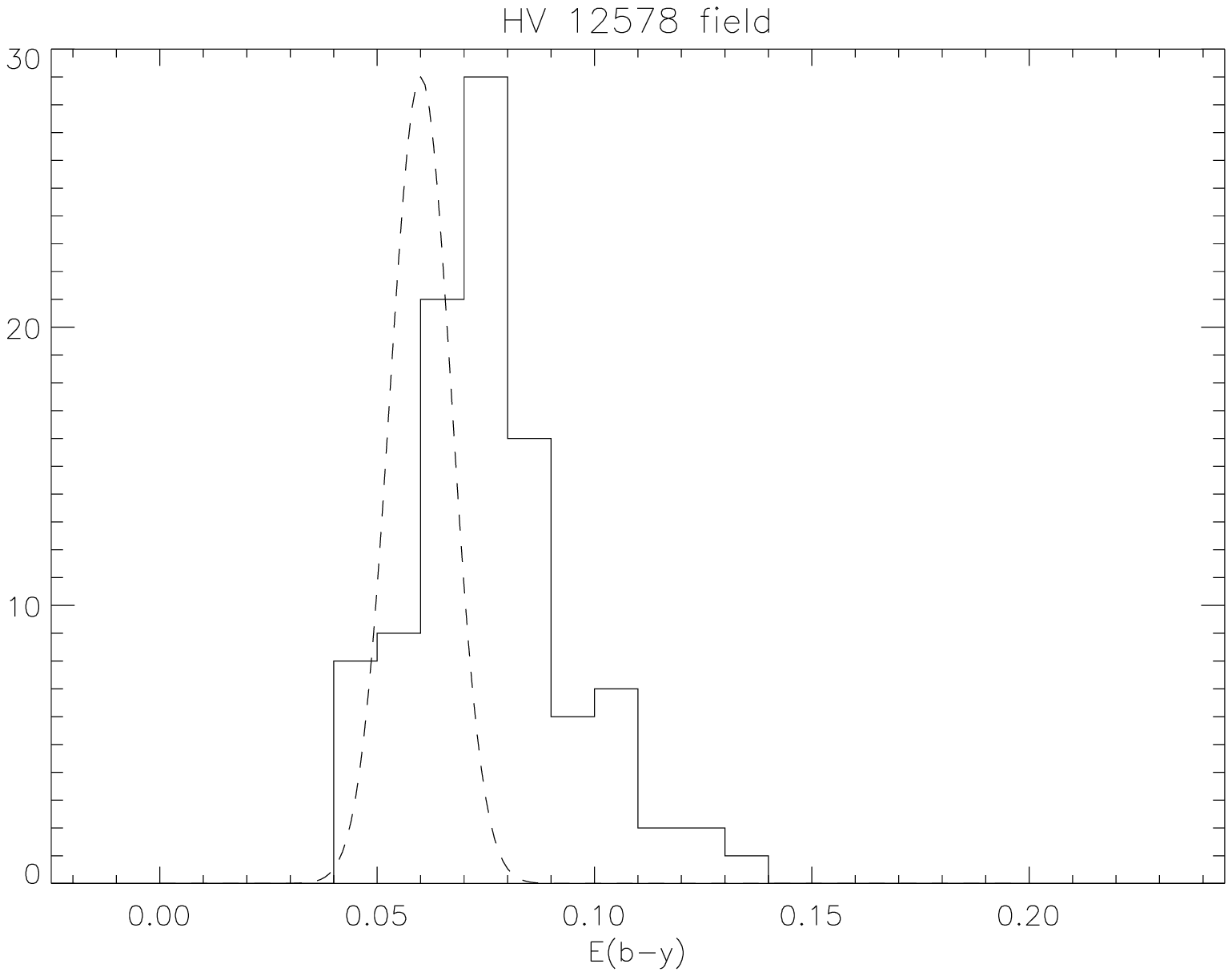}
\epsfxsize=8cm
\epsfbox{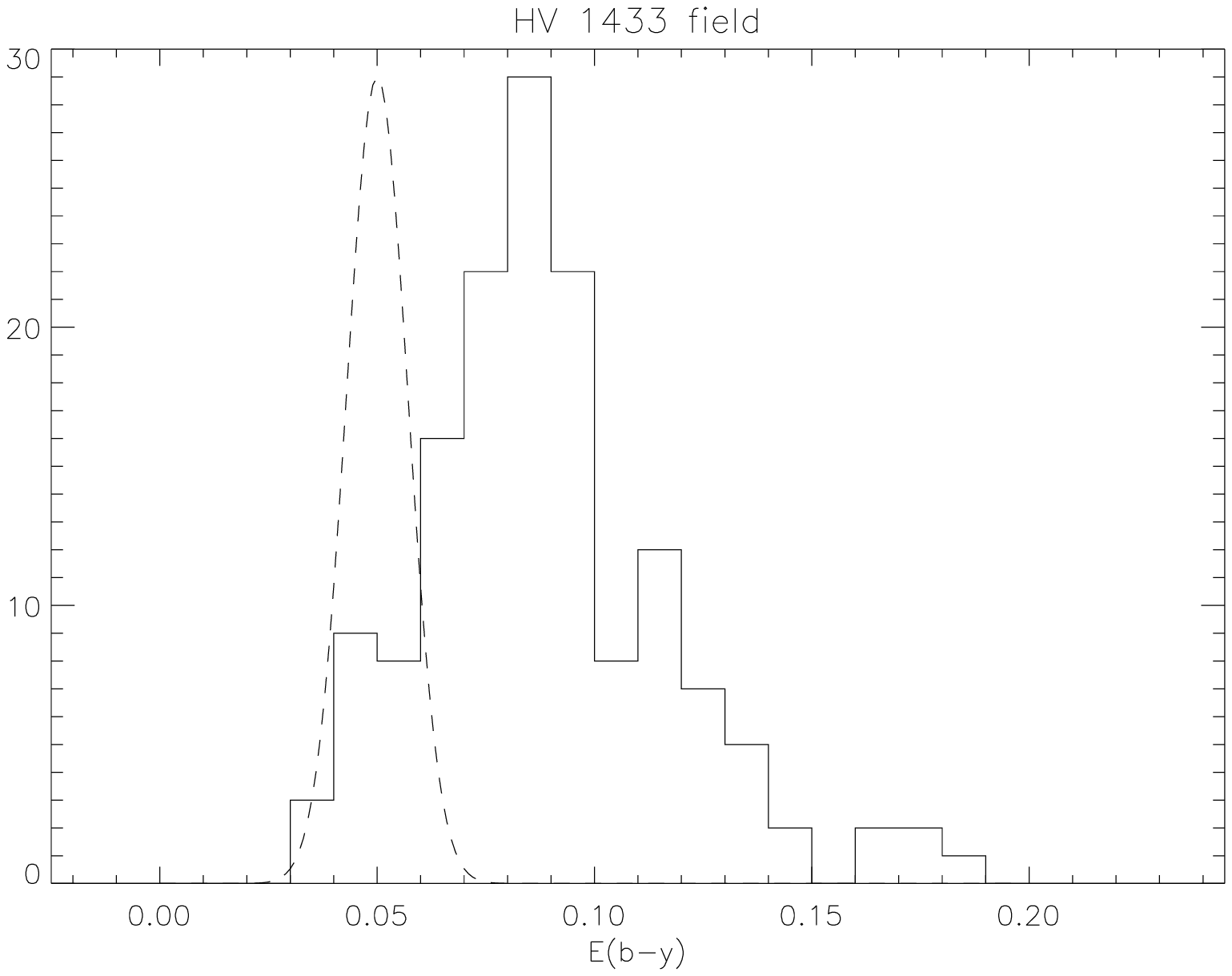}
\epsfxsize=8cm
\epsfbox{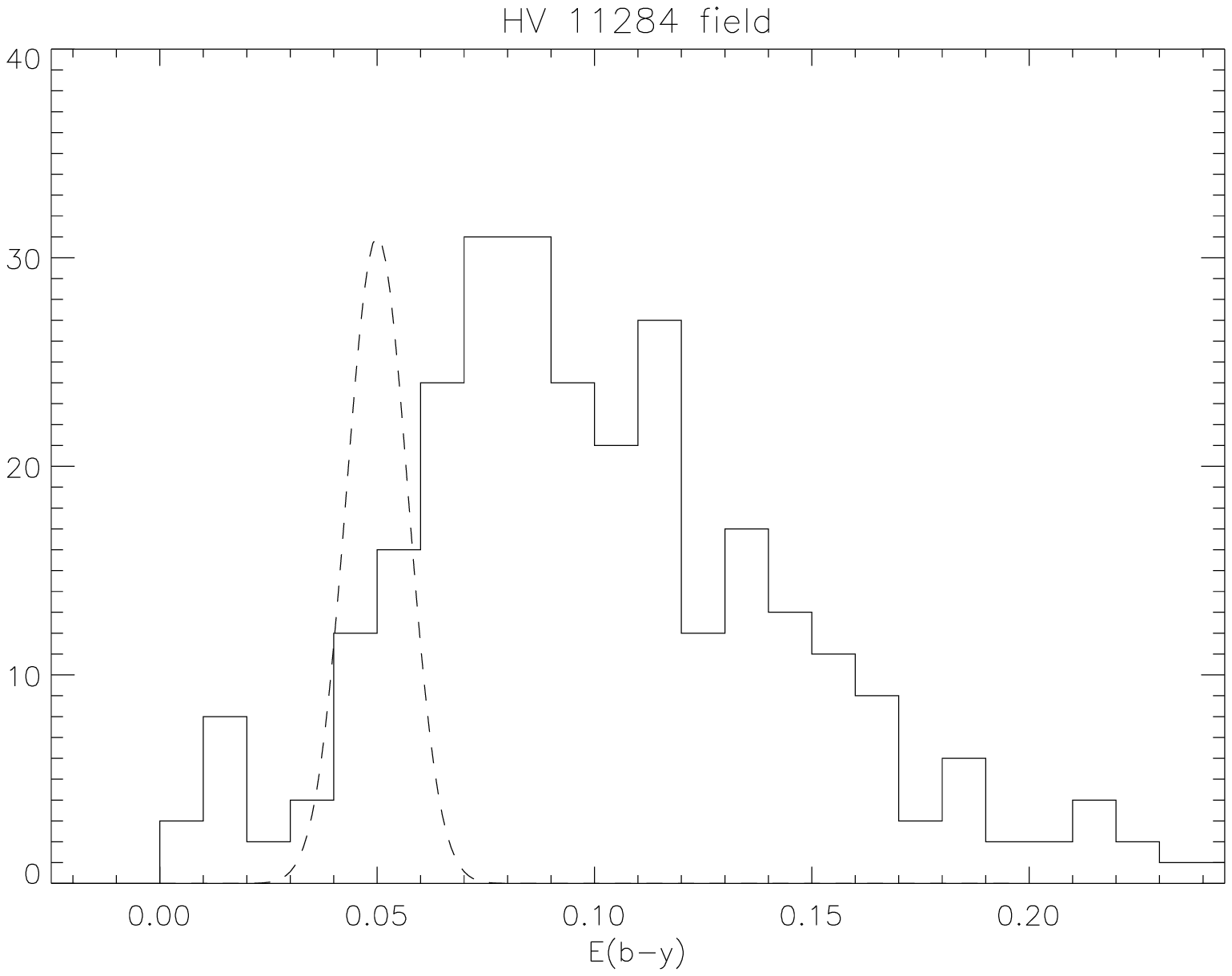}
\caption[]{\label{fig:mcred}Reddening histograms (B stars) for the four fields;
see text for details.}

\end{figure}

  The $(b-y),c_1$ diagrams for B stars in each of the four fields are
shown in Fig.~\ref{fig:byc1} together with a line 
corresponding to the
standard relation of Crawford (\cite{craw1978}). The arrow shows the direction
of the reddening vector. Only stars with 
a photometric error in $(b-y)$ of less than 0.015 and less than 0.05 in 
$c_1$ according to DAOPHOT were included in the analysis. 
Stars with a DAOPHOT $\chi$ estimate larger than 2 in $uvby$
were also rejected regardless of the estimated error. The left-most 
boundary of the data points is quite well-defined, but offset with respect
to the $(b-y)_0, c_0$ standard relation (due to foreground reddening), while 
the right-most boundary is more diffuse. In the SMC HV11284 field we note 
the presence of a few stars with apparently very low reddenings. 

  Once again, we used the Padua isochrones combined with Kurucz atmospheres
to test if the Crawford (\cite{craw1978}) relation applies also to the
metal-poorer LMC/SMC stars. The theoretical $(b-y),c_1$ diagrams
for stars with $1<M_V<-4$ are shown in Fig.~\ref{fig:byc1_synt} for four
different metallicities together
with the Crawford relation. Like in Fig.~\ref{fig:m1c1_synt}, five
different isochrones corresponding to ages between $10^7$ and $10^9$
years are included but no random errors are added in Fig.~\ref{fig:byc1_synt}.
The models and the empirical relation generally agree well for all 
metallicities and there are no evident trends with metallicity that
would affect the reddening determinations, so we conclude that reddenings 
derived from the $(b-y),c_1$ diagram are reliable.

\begin{table}
\caption[]{
\label{tab:eby} Basic reddening characteristics for
the observed fields. Columns 2 and 3 give the minimum and average reddening
for each field according to our investigation based on B stars.}
\begin{flushleft}
\begin{tabular}{lcc} \hline
Field    & $E(b-y)$, min. & $E(b-y)$, avg. \\ \hline
\multicolumn{1}{l}{LMC:} & & \\
HV982   &  0.060 $\pm$ 0.015       &  0.100        \\ 
HV12578 &  0.060 $\pm$ 0.015       &  0.075        \\ 
\multicolumn{1}{l}{SMC:} & & \\
HV1433  &  0.050 $\pm$ 0.015       &  0.089       \\
HV11284 &  0.050  $\pm$ 0.015      &  0.100       \\ \hline
\end{tabular}
\end{flushleft}
\end{table}

  The reddening distributions derived from the data in Fig.~\ref{fig:byc1} 
as described in the previous section are shown in Fig. \ref{fig:mcred}. 
The dashed Gaussians represent the average observational scatter.
The lower limit of the reddening distribution is interpreted 
as being caused by Galactic foreground extinction. We estimate that this
 amounts to $E(b-y) = 0.060 \pm 0.015$ in both of the LMC fields and 
$E(b-y) = 0.050 \pm 0.015$ in the SMC fields, corresponding to 
$E(B-V) = 0.085 \pm 0.02$ and $E(B-V) = 0.07 \pm 0.02$, respectively. The 
estimated errors in these reddenings are largely due to zero-point errors 
in the $b$ and $y$ photometry, while errors in $c_1$ do not affect
the results significantly.

For the LMC fields, we can compare the foreground reddening determinations 
to the reddening map by Oestreicher et al. (\cite{oest1995}), which predicts 
foreground reddenings of $E(B-V) = 0.045$ and $E(B-V) = 0.050$ 
in the HV12578 and HV982 fields. Thus, our estimated foreground reddenings are 
somewhat higher and agree well with the typical value of $E(B-V) = 0.075$ reported by
Schlegel et al. (\cite{schlegel1998}). However, their typical SMC value of
$E(B-V) = 0.037$ is significantly lower than our results.

\subsection{The correlation of reddening with position.}

  The width of the histograms representing the reddening distributions
cannot be accounted for by observational uncertainties and must, 
consequently, represent real scatter in the reddenings of LMC/SMC
B stars.  This could either mean that the foreground 
reddening varies by as much as 0.1 in $E(B-V)$ within the fields,
or that the B stars are subject to different amounts of reddening 
from the interstellar medium in the Clouds themselves, being located at
different optical depths.
In the first case, we would expect that reddenings 
are strongly correlated with the position in the image, whereas such a
correlation would be weaker or absent if the variations are caused 
by depth effects.  Of course, any intermediate scenario is in 
principle possible.

If reddening is correlated with position then the difference
between the reddenings of two neighbouring stars will, on the average,
be smaller than that of two widely separated stars.  
In order to investigate this effect, we used our measurements of
individual B star reddenings to calculate the r.m.s. reddening difference
between two stars as a function of their separation in the image.
We denote the r.m.s reddening difference between the reddening of a
star and other stars at distances $[R \ldots R+\Delta R[$
from it $C(R,\Delta R)$ where $\Delta R$ is the bin size.  We found 
$\Delta R = 10$ pixels to be a reasonable bin size.

If reddening is correlated with 
position on scales smaller than some characteristic limit, $C(R,\Delta R)$ 
will decrease as $R$ becomes smaller than the characteristic limit, which 
could for example depend on the typical size of interstellar clouds.
$\Delta R$ is not a very critical parameter. It should be chosen sufficiently 
large that a reasonable number of stars will be located within an annulus
of inner radius $R$ and outer radius $R+\Delta R$. On the other hand, it
should not be larger than the typical size of the structure we are looking
for.

\begin{figure}
\epsfxsize=85mm
\epsfbox{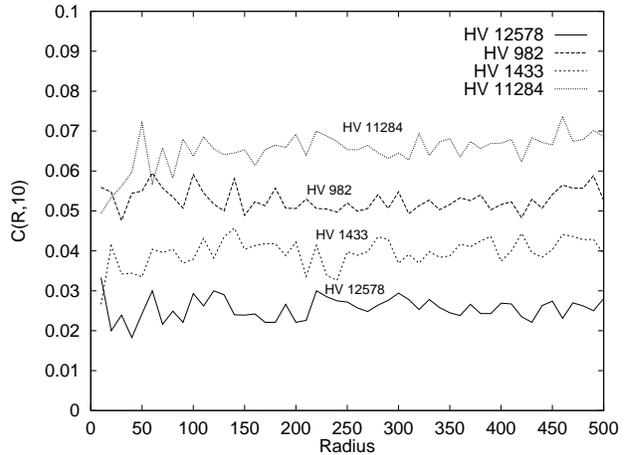}
\caption[]{\label{fig:rms_eby}The quantity $C(R,\Delta R)$, denoting
the rms $E(b-y)$ difference between B stars located in the distance interval
$[R\ldots\Delta R[$ from each other.
  $R = 100$ (pixels) corresponds to 38$\arcsec$.}
\end{figure}

  Figure \ref{fig:rms_eby} shows $C(R,\Delta R)$ for each field, with
$\Delta R$ set to 10 pixels or roughly 4 arcseconds.  There is no
significant decrease in $C$ even for quite small $R$, with a hint of such
a decrease only in the SMC fields (HV1433, HV11284). 
At the same time, Fig.~\ref{fig:rms_eby}
also gives for each field an indication of how large the typical error
will be if just the average reddening of the field is used as an
estimate of the reddening for a given star. This varies from one 
field to another, from $E(b-y)$ = 0.025 in the HV12578 field to 0.07 in 
the HV11284 field, and it is clear that the error is in no case 
negligible. Based on the two fields we have observed in each galaxy,
the contribution from internal reddening in the Clouds appears to decrease
as a function of distance to the centre.

\subsection{Discussion of reddenings}

  The previous section has shown a lack of correlation between reddening and
position in the fields, even on the smallest scales on which this can be
investigated by means of the present Str{\"o}mgren photometry, i.e. a few arc
seconds (for LMC and SMC, 1 pc corresponds to about $4\arcsec$ and
$3\farcs5$, respectively).

 Many stars have reddenings much larger than the Galactic foreground
contribution, in some fields up to $E(b-y) = 0.15$ ($E(B-V) = 0.21$) or so. 
Because the reddening is not correlated with the position in 
the field, we conclude that the variations are most likely due to the stars 
being located at different depths in the Magellanic Clouds. Therefore, the 
difference between the maximum and minimum reddening in each field presumably 
represents the total reddening when looking through the LMC or SMC at the 
corresponding position.

It is of interest to compare our results with the investigation by 
Oestreicher \& Schmidt-Kaler (\cite{oest1996}). Their reddening 
map shows a lack of stars with high reddenings in the neighbourhood of 
the HV12578 field, in 
agreement with our results, whereas a large number of stars with 
relatively high reddenings are found near the 30 Dor region, again in 
agreement with our results. Oestreicher \& Schmidt-Kaler (\cite{oest1996}) 
did not study the Small Magellanic Cloud.  Olsen (\cite{olsen99}) only 
found strong differential reddening in one out of four LMC fields observed
with the HST, centered on the cluster NGC~1916 near the centre of the LMC 
bar.

  As we shall see in the following section, the fact that average reddenings 
are only accurate to within several hundreths of a magnitude is potentially a 
serious problem for photometric studies of stars where reddenings cannot 
be directly determined, such as metallicity studies of GK giants in the 
Str{\"o}mgren system (Hilker et al.~\cite{hilk1995};
Grebel \& Richtler \cite{greb1992}; Dirsch et al.~\cite{dirsch2000}).

\section{Metallicities}

\subsection{Metallicities for GK giants}

\begin{table*}
\caption[]{\label{tab:metstat}Metallicity data for the four fields.
Numbers in parantheses indicate the values when only stars with
[Fe/H]$<$0.0 are included. N is the number of GK giants used.
}
\begin{flushleft}
\begin{tabular}{lrrlll} \hline
Field ID & N & & [Fe/H](Mean)  & [Fe/H](Med)   & [Fe/H](Mode)  \\ \hline
\multicolumn{1}{l}{LMC:}\\
HV982   & 121& (92)& $-0.40$ ($-0.65$) & $-0.32$ ($-0.54$) & $-0.15$ ($-0.33$) 
\\
HV12578 & 102& (95)& $-0.47$ ($-0.53$) & $-0.42$ ($-0.44$) & $-0.31$ ($-0.27$) 
\\
\multicolumn{1}{l}{SMC:}\\
HV1433  & 167&(155)& $-0.68$ ($-0.76$) & $-0.64$ ($-0.66$) & $-0.56$ ($-0.46$) 
\\
HV11284 &  86& (74)& $-0.46$ ($-0.60$) & $-0.45$ ($-0.54$) & $-0.42$ ($-0.42$) 
\\
\hline
\end{tabular}
\end{flushleft}
\end{table*}

For GK giants it is appropriate to define the metallicity calibration 
directly in terms of $(b-y)$ and $m_1$ instead of using the $\delta m_1$ 
notation for F dwarfs of e.g. Crawford (\cite{craw1975}).  We use the 
calibration by Hilker (\cite{hilker99}), 
\begin{equation}
  \mbox{[Fe/H]} = \frac{m_0 \, + a_1 (b-y)_0 \, + a_2}{a_4 \, + a_3 (b-y)_0}
    \label{eq:FeHcali}
\end{equation}
with $a_1 = -1.277\pm0.050$, $a_2 = 0.331\pm0.035$, 
$a_3 = 0.324\pm0.035$ and $a_4 = -0.032\pm0.025$, valid
for $-2.0 < \mbox{[Fe/H]} < 0.0$. When plotted in the $(b-y)_0, m_0$ diagram 
(i.e.  after de-reddening), GK giant stars of the same metallicity will fall 
along a straight line, independently of luminosity.  
Compared to an earlier calibration
of [Fe/H] as a function of $(b-y)_0$ and $m_0$ by Grebel \& Richtler 
(\cite{greb1992}), the Hilker (\cite{hilker99}) calibration yields
somewhat lower metallicities, especially at low metallicities. The difference
amounts to about 0.1 dex for [Fe/H]$\sim-0.5$ and increases to $\sim0.25$ 
dex for [Fe/H]$\sim-1.5$.

  For a ``typical'' GK giant with $(b-y)_0 = 0.75$ and $m_0 = 0.5$, errors 
in $v$, $b$ and $y$ of 0.01 magnitudes correspond to a total error in the 
derived metallicity of about 0.16 dex, with the
most significant contribution to the total error arising from the $b$ band 
error. Similarly, an error in the estimated reddening $E(b-y)$ of 0.01 mag 
translates to about 0.05 dex in the derived [Fe/H]. Hence, the derived
metallicities are quite sensitive to errors in the measurements as well
as in the assumed reddenings, and great care must be taken to avoid
systematic errors which may shift the observed metallicity distributions
by significant amounts.

\begin{figure}
\epsfxsize=8cm
\epsfbox{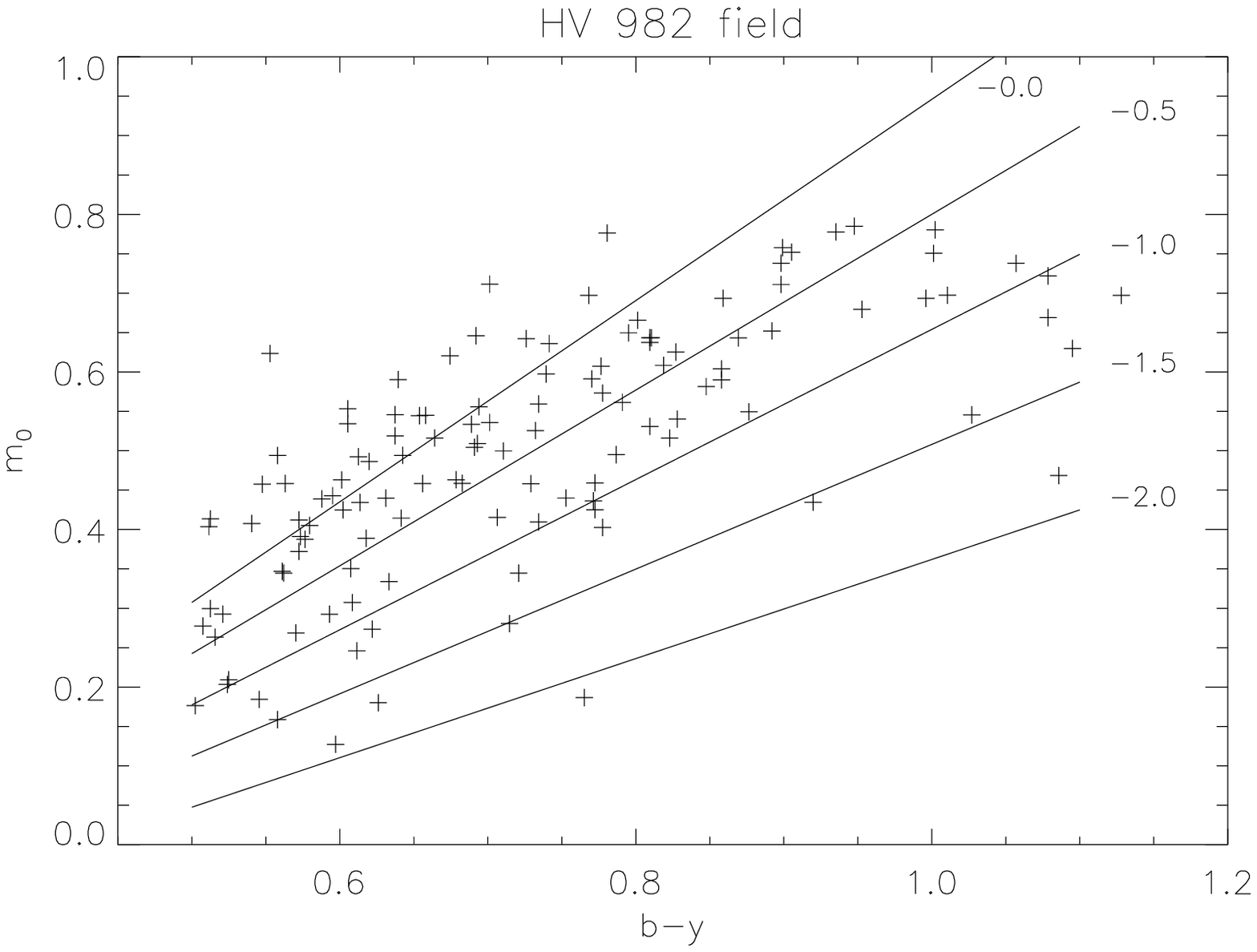}
\epsfxsize=8cm
\epsfbox{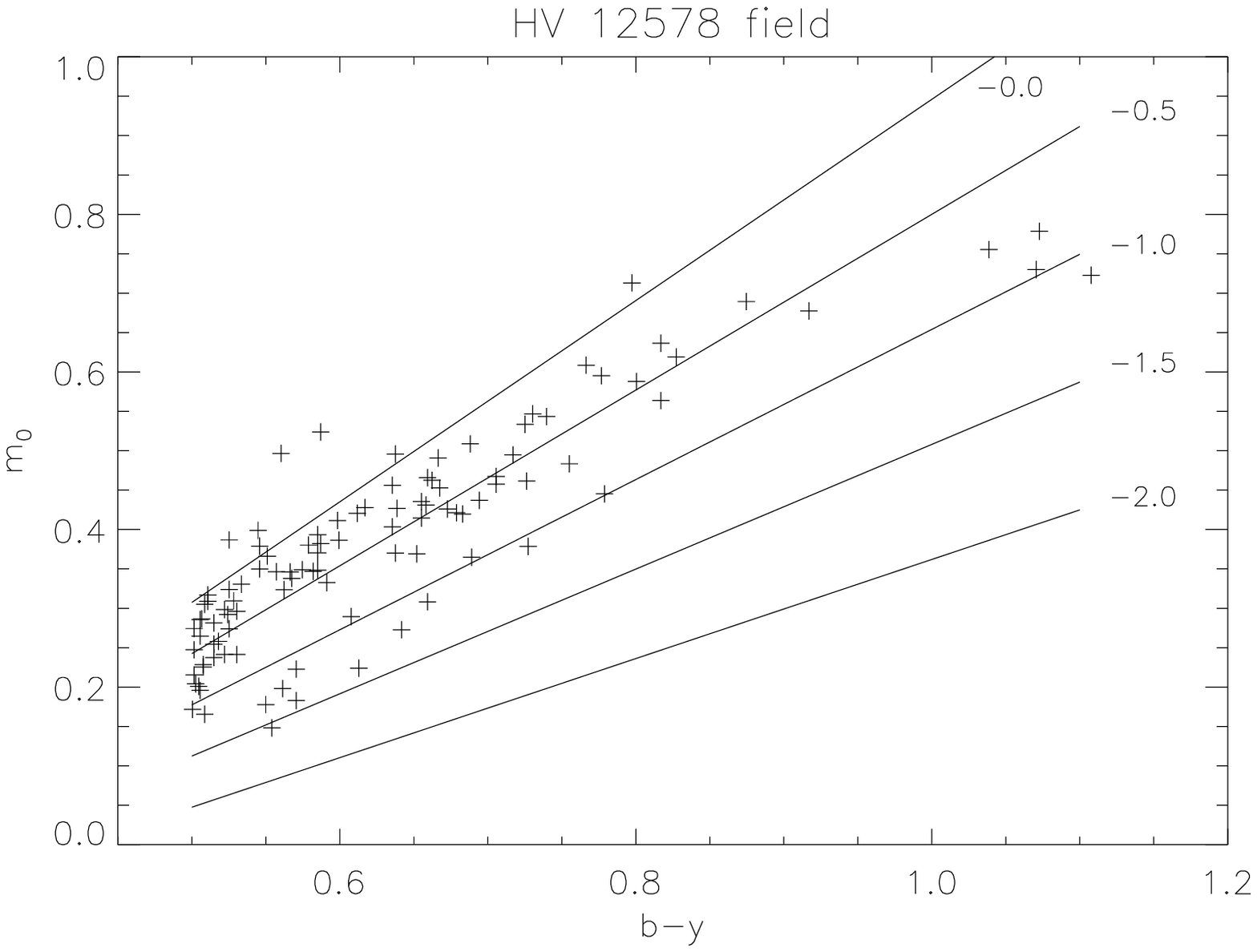}
\epsfxsize=8cm
\epsfbox{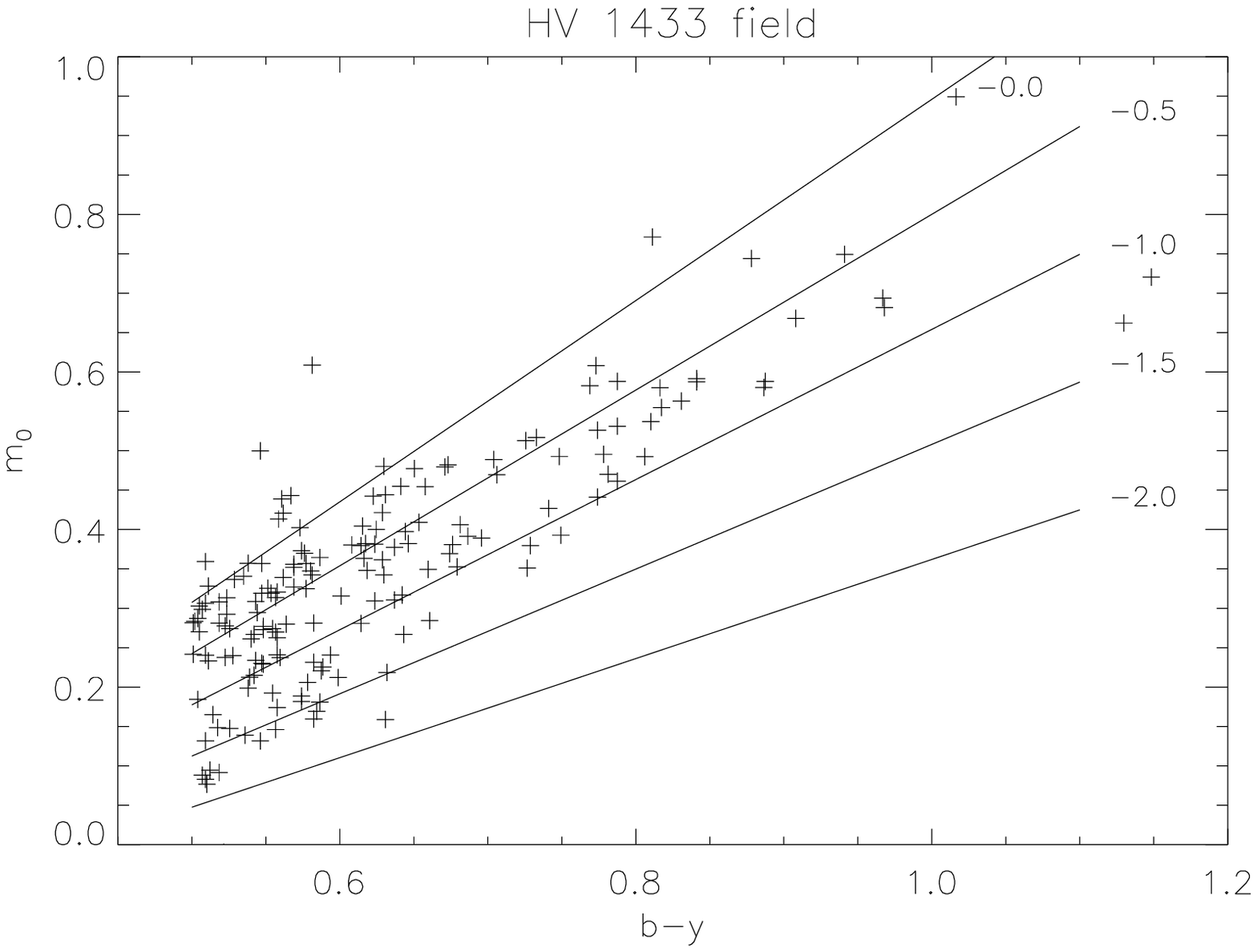}
\epsfxsize=8cm
\epsfbox{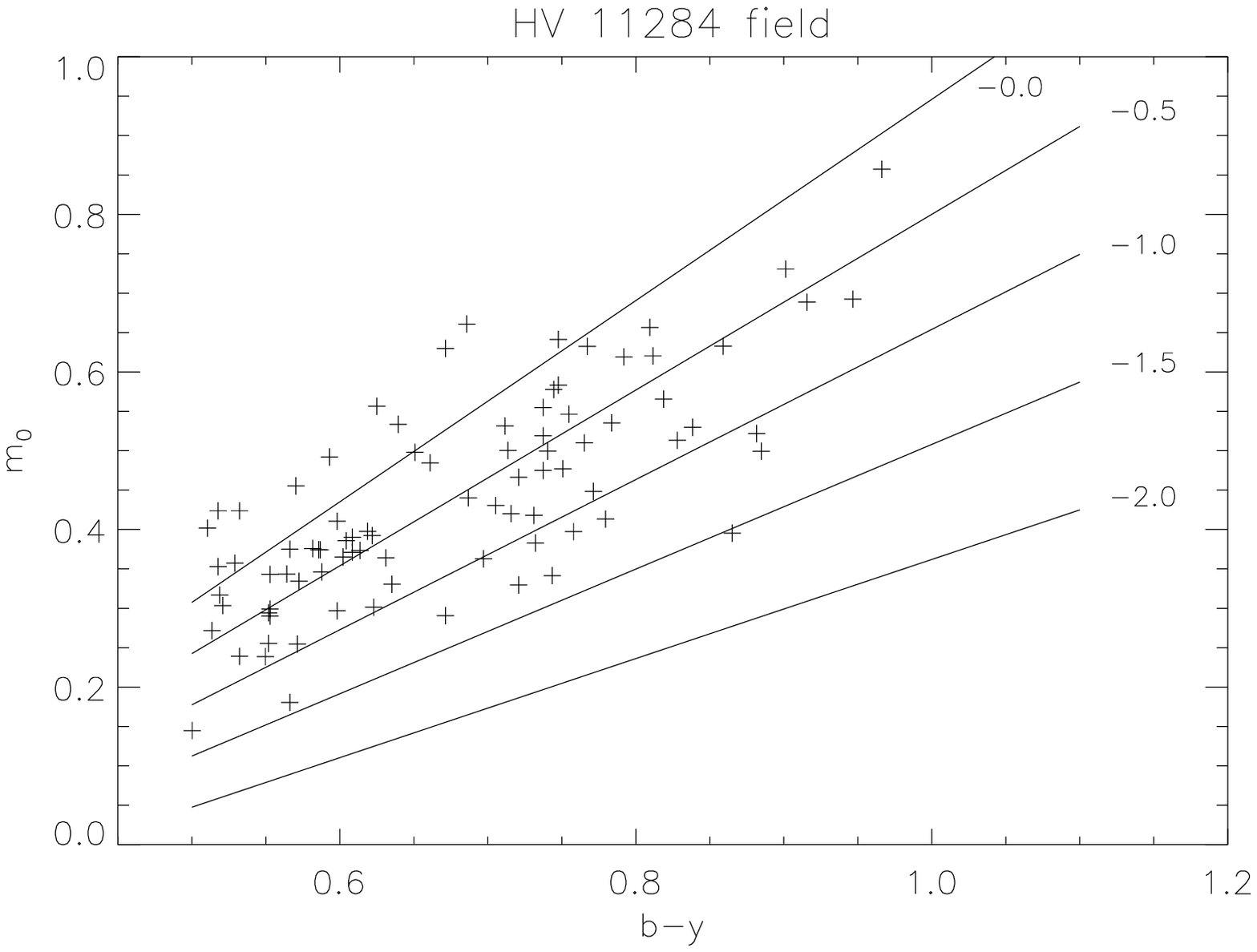}
\caption[]{\label{fig:bym1}$(b-y)_0, m_0$ diagrams for the four fields.
The straight lines represent constant metallicities according to
Eq.~\ref{eq:FeHcali}.
}
\end{figure}

\begin{figure}
\epsfxsize=8cm
\epsfbox{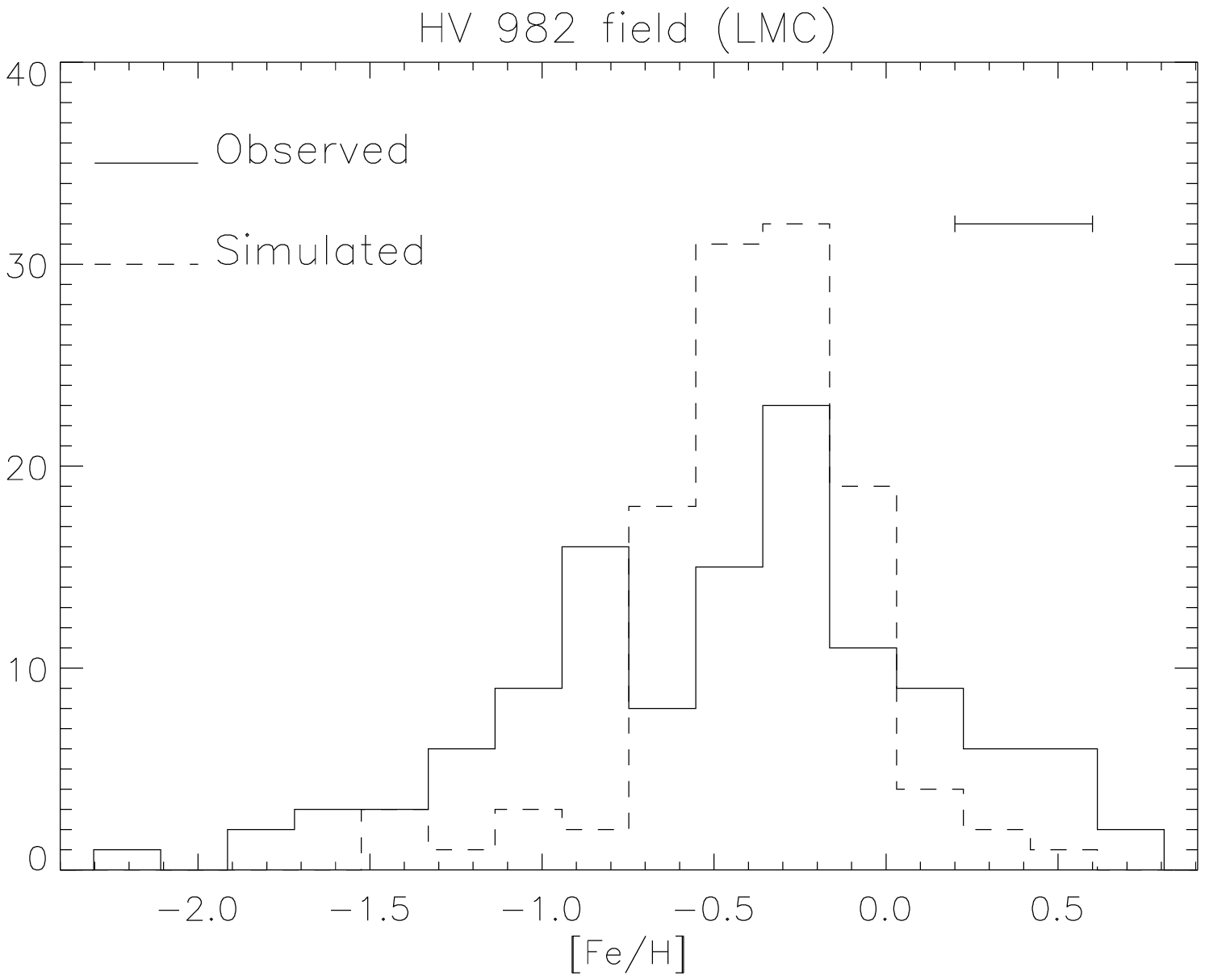}
\epsfxsize=8cm
\epsfbox{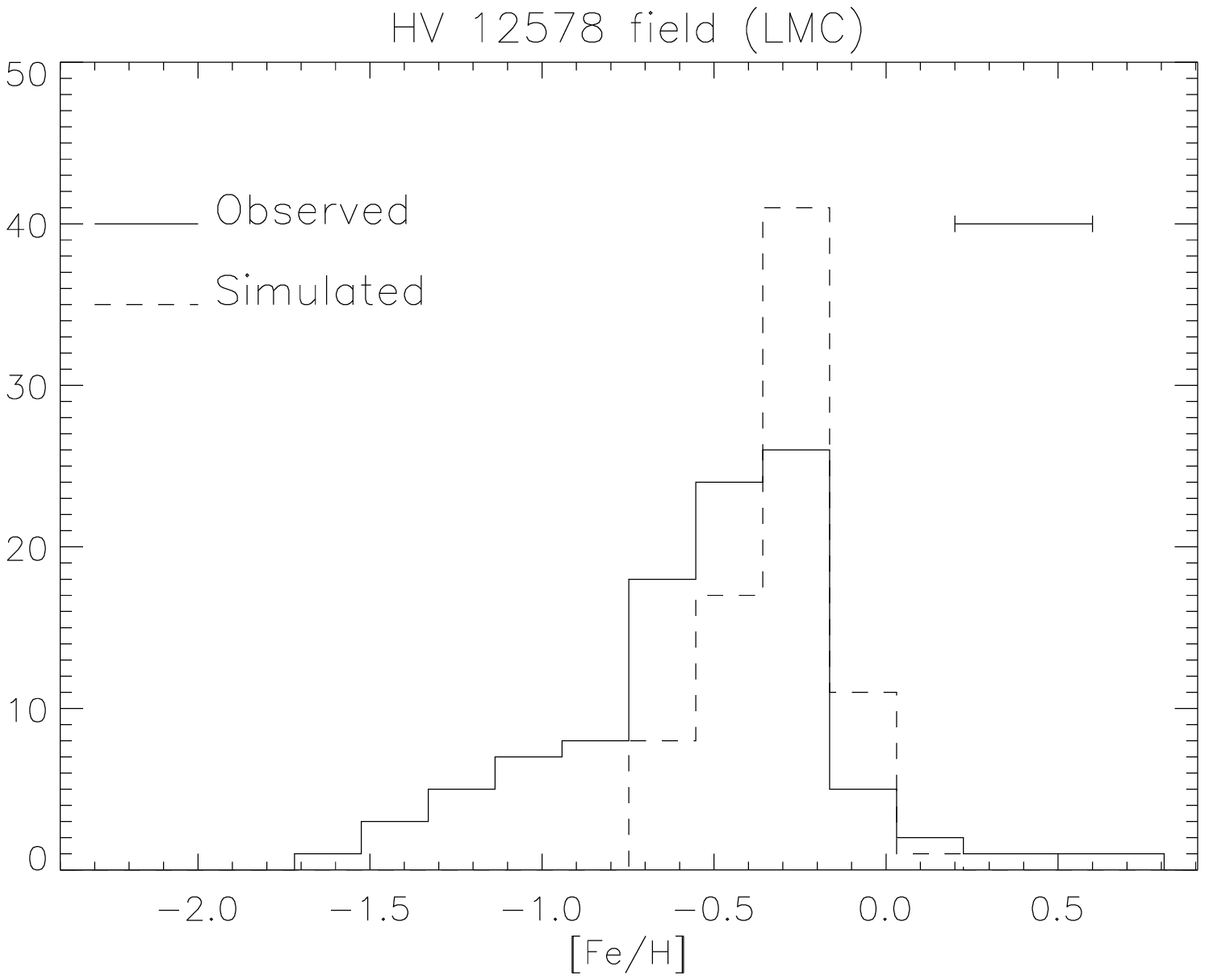}
\epsfxsize=8cm
\epsfbox{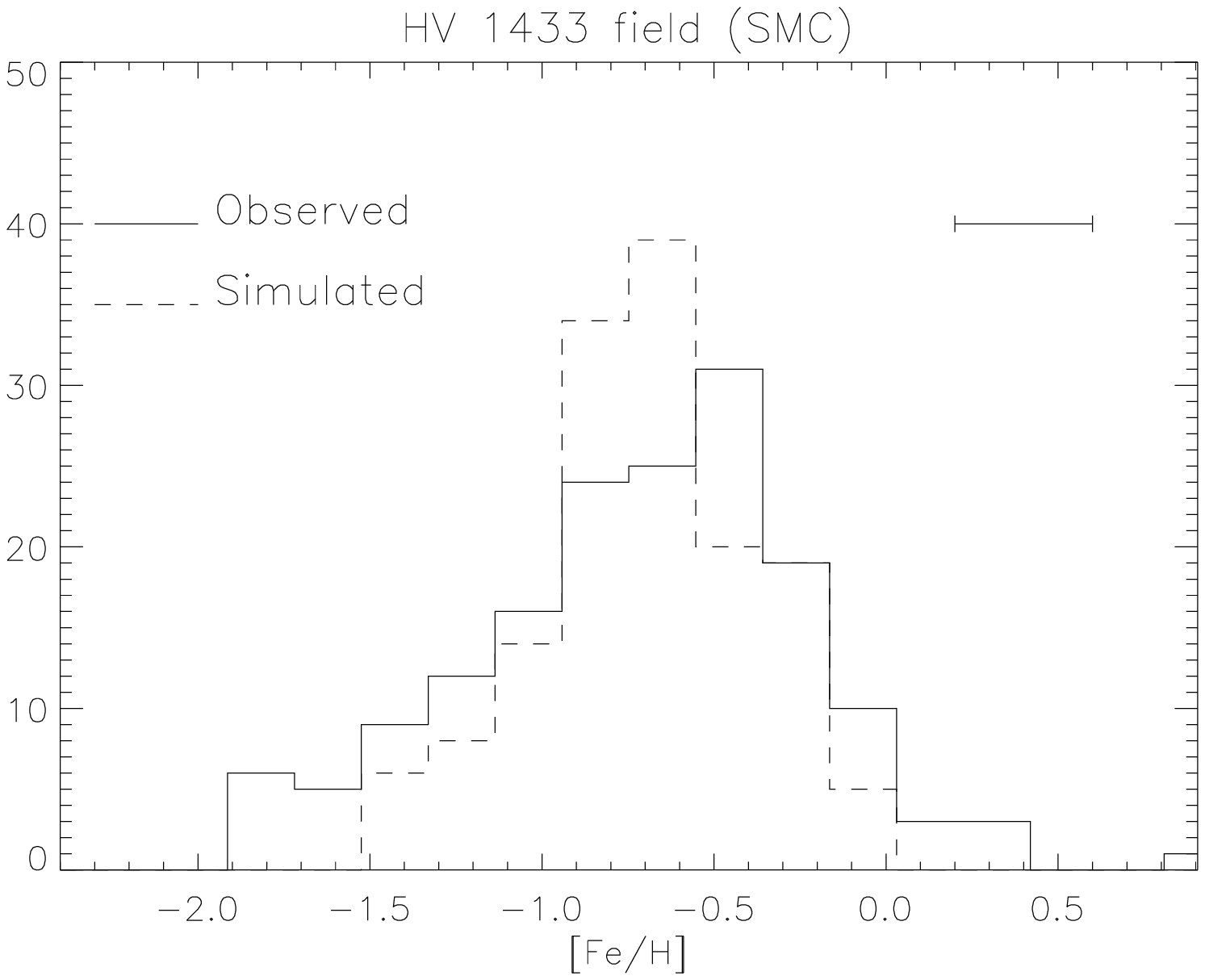}
\epsfxsize=8cm
\epsfbox{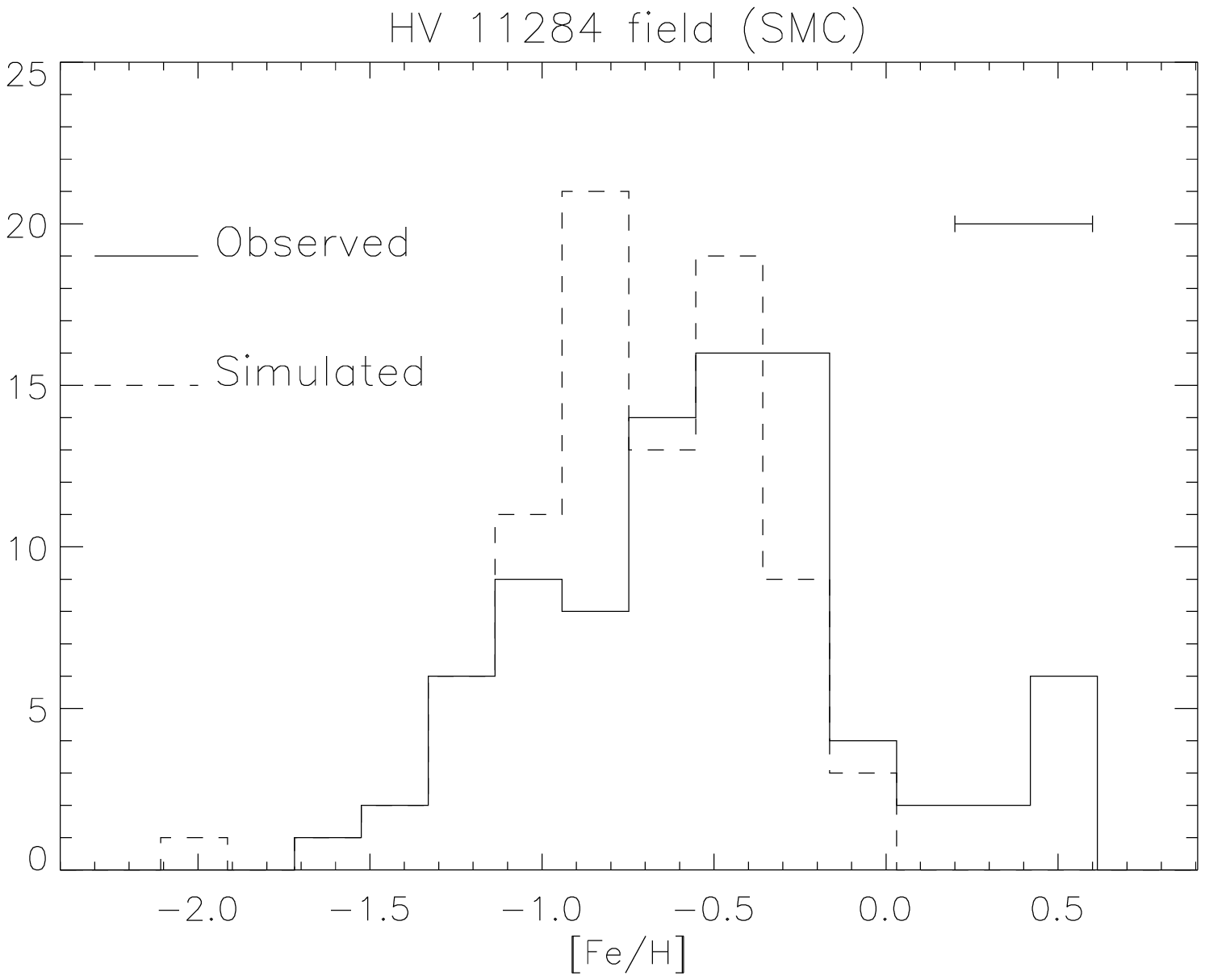}
\caption[]{\label{fig:metal}Metallicities for the four fields; see text.
The error bars refer to systematic errors due to zero-point errors in
calibration of the photometry.
}
\end{figure}

  In this section we analyze the metallicity distributions derived for
GK giants using the calibration given by Eq. (\ref{eq:FeHcali}).  Following 
Hilker (\cite{hilker99}) we have used the calibration for stars with 
$0.5 < (b-y) < 1.1$. Additional selection criteria 
based on DAOPHOT parameters were err$(m_1) < 0.02$ and $\chi(y,b,v) < 2.0$. 
This resulted in about $100-200$ stars for metallicity analysis in each field
(see Table \ref{tab:metstat}).
  
Figure~\ref{fig:bym1} shows the $(b-y)_0, m_0$ diagrams for the four fields
and the derived metallicity distributions are in Figure~\ref{fig:metal}
(solid lines). The average reddenings given in Table \ref{tab:eby} have
been used.  Fig.~\ref{fig:metal} shows a significant
scatter in the derived metallicities, from [Fe/H] = $-2.0$ up to around
\mbox{[Fe/H] = 0}.  In some of the fields the derived metallicity 
distributions include a number of stars with [Fe/H] $>$ 0. However, as shown 
for the Milky Way globular cluster 47 Tuc by Dickens et al. 
(\cite{dick1979}), stars with peculiar CNO abundances can mimic stars with 
a generally high metallicity.  The effect of CN anomaly on metallicities 
derived from Str{\"o}mgren photometry has also been illustrated by Richter 
et al.~(\cite{richter99}).  Furthermore, the $(b-y)_0, m_0$ metallicity 
calibration is valid only for subsolar metallicities.  We therefore can 
not conclude with certainty from our data that stars with truly high 
metallicities exist in the LMC or SMC, while the possibility remains open. 
Spectroscopic studies will be needed in order to answer this question 
definitively.

   Average, median and mode statistics for the metallicity distributions
are listed in Table~\ref{tab:metstat}, with numbers in parantheses based on 
[Fe/H] values less than 0. The SMC fields generally come out more metal poor
than the LMC fields, although the SMC HV11284 field appears to be nearly
as metal-rich as the LMC HV12578 field. However, the uncertainty on the
photometric zero-points translates to roughly 0.2 dex in [Fe/H], so within
the error limits the metallicities are consistent with the results from 
spectroscopic studies of F and G supergiants, \mbox{[Fe/H] = $-0.3 \pm 0.2$} 
for the LMC and \mbox{[Fe/H] = $-0.65 \pm 0.2$} for the SMC 
(Westerlund \cite{west1997}). Our metallicities for the LMC fields are
also consistent with those obtained by Dirsch et 
al.~(\cite{dirsch2000}) who quote [Fe/H] values in the range 
$-0.71\pm0.23$ to $-0.52\pm0.21$ dex for young LMC clusters. 

\subsection{Investigating the effect of reddening}
\label{sec:ieor}

  With the knowledge of reddenings obtained from B stars, it is 
possible to estimate how much of the apparent scatter in metallicity 
seen in Fig.~\ref{fig:metal} may actually be attributed to reddening 
variations.

  In order to investigate how reddening variations affect the derived
metallicity distributions, we carried out the following experiment: First, 
a set of $(b-y), m_1$ data pairs were generated, corresponding to one single 
metallicity (the canonical values were used for this experiment). This was 
accomplished by using the list of observed $(b-y)$ indices and then generating 
the $m_1$ indices from the calibration equation. When plotted in 
the $(b-y), m_1$ diagram these points would then per definition fall along 
straight lines. Next, for each $(b-y),m_1$ data pair the reddening of a 
randomly selected B stars was added. The ``metallicities'' were then 
determined for these synthetic data using an {\it average} reddening, in the
same way as for the real GK giants. This 
lead to the histograms drawn with dashed lines in Fig. \ref{fig:metal}. 

  The peaks of the simulated metallicity distributions are in quite 
good agreement with those of the actual observed distributions, at least to 
within the uncertainty arising from zero-point errors in the photometry.
The scatter in the observed metallicity distributions remains somewhat larger 
than that of the simulated ones, and in particular, we note the presence 
of what might be interpreted as a metal-poor population, with metallicities 
extending down to [Fe/H] $\approx -2.0$. In order to quantify to what
extent some of the scatter in the metallicity distributions may be
intrinsic, we compared the observed and simulated distributions
using an F--test (Press et al.~\cite{pre92}). For all the fields,
the hypothesis that the variances of the two distributions are
similar is rejected at a very high ($>99$\%) confidence level.
This remains true even if the analysis is restricted to data with
[Fe/H]$<0$, except for the HV11284 field where no statistically significant
difference is now found between the variances of the observed and
simulated metallicity distributions.

  It is instructive to consider what happens when one tries to compare e.g. 
the metallicity of cluster stars and the metallicity of surrounding field star 
populations: The cluster stars will all be located at the same depth in the 
LMC or SMC, so they will all be affected by the same amount of interstellar 
absorption. On the other hand, the field stars will be randomly 
distributed radially, and therefore their reddenings will vary accordingly. 
When metallicities are derived using Str{\"o}mgren photometry one will indeed 
be able to confirm that the cluster stars all have the same metallicity, 
seemingly proving that the Str{\"o}mgren photometry ``works'', while the field 
stars will seem to occupy a wide range in metallicity. However, a significant
amount of the scatter in the metallicities derived for the field stars may not 
be real, but is instead due to differences in the reddening from star to star.
Any observed difference between the average field star metallicity and the 
cluster metallicity may be partly real, but will also depend on the amount
of reddening internally in the SMC or LMC to which the cluster is subject.

  For LMC/SMC clusters that are sufficiently young for early-type stars to be 
present it may be worthwhile to consider including $u$ band observations
in future Str{\"o}mgren photometry so that reddenings can be determined.   
With the new generation of UV-sensitive CCD chips this will not be very 
costly in terms of observing time.

\subsection{Ages}

Independent age determinations for the metal-poor stars in our sample could
provide insight into the age-metallicity relation for field stars in the 
Magellanic Clouds. Such a relation is relatively well established for
{\it clusters} (e.g. van den Bergh \cite{van1991}), but the situation for
field stars is more uncertain due to the inherent difficulties in obtaining
independent metallicities and ages. Here we will not attempt to derive
age information for GK giants, but we note that Dirsch et 
al.~(\cite{dirsch2000}) 
attempted to determine an age-metallicity relation and the star formation 
history of both red giant field stars and clusters in their six LMC fields, 
using Str{\"o}mgren $vby$ photometry. They found evidence for an increase 
in the star formation rate $\sim3$ Gyr ago, along with a rapid enrichment. 
However,
the results remain uncertain because of possible CN anomalies in the LMC
GK giants and the problems discussed in Sect.~\ref{sec:ieor}.

\section{Summary and conclusions}

  Based on extensive Str{\"o}mgren $uvby$ photometry 
we have investigated four $6\arcmin\times4\farcm5$ CCD fields, 
two in the LMC and two in the SMC.
Because $u$-band observations were included it has 
been possible to study reddening variations on very small ($\sim1$ pc) scales.

  From B stars we find that reddenings vary by several hundreths of a 
magnitude in $E(B-V)$ within a CCD field. The reddening variations are random 
rather than smooth, so unless the reddening for a particular 
object is directly measured it will be uncertain by at least 0.035 mag
in $E(B-V)$ (HV~12578 field) and up to 0.10 mag (HV~11284 field). This 
translates directly into metallicity errors for red giants of 0.15 dex and 
0.45 dex respectively, unless individual reddenings are available.
We thus suggest that the $u$ filter be included in LMC/SMC Str{\"o}mgren
photometry whenever possible. In particular, metallicity studies of
young star clusters with a suitable number of B stars for reddening 
determinations would benefit strongly from the inclusion of $u-$band data.

  Metallicities have been derived for GK giants. Within the $\sim$0.2 dex
uncertainties, the average metallicities are largely consistent with 
those derived from spectroscopic studies of F and G supergiants,
about \mbox{[Fe/H] = $-0.3$} for the LMC and 
\mbox{[Fe/H] = $-0.65$} for the SMC (Westerlund \cite{west1997}), 
but we also see indications that a smaller number of more metal-poor 
stars with metallicities down to \mbox{[Fe/H] = $-2.0$} are present in 
each of the fields. There are also hints of stars with higher 
metallicities, but this might be an effect of chemical peculiarities such as 
enhanced CNO abundances which are known to affect metallicities derived
for red giants by means of Str{\"o}mgren photometry in Galactic globular
clusters (Dickens et al. \cite{dick1979}; Richter et al. \cite{richter99}). 
Further follow-up spectroscopic studies are needed.

  An unexpected byproduct of this study is the peculiar location of some
early-type stars in the $[m_1],[c_1]$ diagram. Comparison with stellar 
models shows that this may be partly due to the lower-metallicity 
environments in the LMC and SMC, combined with the effect of random 
photometric errors. However, spectroscopy of some of the stars with 
peculiar indices would be highly desirable in order to check if the 
early-type stars in the LMC/SMC show other physical differences with 
respect to stars in the Solar neighbourhood.

\begin{acknowledgements}
  ESO is gratefully acknowledged for granting a 2-months studentship
to SSL in 1995, during which parts of the data reduction were carried out. 
This research was supported by the Danish Natural Science Research Council
through research grants and through its Centre for Ground-Based Observational 
Astronomy. We thank the anonymous referee for useful suggestions and 
comments which helped improve the paper.

\end{acknowledgements}

\end{document}